\documentclass[twocolumn]{aastex63}
\usepackage{lineno}
\usepackage{inputenc}
\usepackage{booktabs}
\usepackage{latexsym}
\usepackage{graphicx}
\usepackage{amssymb}
\usepackage{longtable}
\usepackage{epsf}
\usepackage{lineno}
\usepackage{float}
\usepackage{comment}
\usepackage[]{natbib}
\usepackage{placeins}
\usepackage{multirow}
\usepackage{amsmath} 
\usepackage{hyperref}
\usepackage{booktabs}

\graphicspath{{./figures/}}

\begin{document}
\title{Hundreds of Low-Mass Active Galaxies in the Galaxy And Mass Assembly (GAMA) Survey}

\author[0000-0002-4587-1905]{Sheyda Salehirad}
\affiliation{eXtreme Gravity Institute, Department of Physics, Montana State University, Bozeman, MT 59717, USA }
\author[0000-0001-7158-614X]{Amy E.\ Reines}
\affiliation{eXtreme Gravity Institute, Department of Physics, Montana State University, Bozeman, MT 59717, USA }
\author[0000-0001-8440-3613]{Mallory Molina}
\affiliation{eXtreme Gravity Institute, Department of Physics, Montana State University, Bozeman, MT 59717, USA }
\affiliation{Department of Physics and Astronomy, University of Utah, 115 South 1400 East, Salt Lake City, UT 84112, USA}

\begin{abstract}

We present an entirely new sample of 388 low-mass galaxies ($M_\star \leq 10^{10} M_\odot$) that have spectroscopic signatures indicating the presence of massive black holes (BHs) in the form of active galactic nuclei (AGNs) or tidal disruption events (TDEs). Of these, 70 have stellar masses in the dwarf galaxy regime with $10^8 \lesssim M_\star/M_\odot \lesssim 10^{9.5}$. We identify the active galaxies by analyzing optical spectra of a parent sample of $\sim$23,000  low-mass emission-line galaxies in the Galaxy and Mass Assembly (GAMA) Survey Data Release 4,
and employing four different diagnostics based on narrow emission line ratios and the detection of high-ionization coronal lines. 
We find that 47 of the 388 low-mass active galaxies exhibit broad H$\alpha$ in their spectra, corresponding to virial BH masses in the range $M_{\rm BH} \sim 10^{5.0-7.7} M_\odot$ with a median BH mass of $\langle M_{\rm BH}\rangle \sim 10^{6.2} M_\odot$. 
Our sample extends to higher redshifts ($z \le 0.3; \langle z \rangle=0.13$) than previous samples of AGNs in low-mass/dwarf galaxies based on Sloan Digital Sky Survey spectroscopy, which can be attributed to the spectroscopic limit of GAMA being $\sim 2$ magnitudes deeper. Moreover, our multi-diagnostic approach has revealed low-mass active galaxies spanning a wide range of properties, from blue star-forming dwarfs to luminous ``miniquasars" powered by low-mass BHs. As such, this work has implications for BH seeding and AGN feedback at low masses.    
\end{abstract}



\section{Introduction}\label{sec:intro}
Supermassive black holes (BHs) are found in the nuclei of almost all massive galaxies \citep[e.g.][]{Kormendy:1995,Kormendy:2013}, however the ``memory'' of BH seeding is erased during merger-driven growth over cosmic time \citep[e.g.][]{Volonteri:2010,Natarajan:2014}. The current proposed seeding models include remnants of Population III stars \citep[e.g.,][]{Bromm:2011}, direct collapse scenarios \citep[e.g.,][]{Loeb:1994,Begelman:2006,Lodato:2006}, and runaway collisions in dense star clusters \citep[e.g.,][]{Portegies:2004,Devecchi:2009,Miller:2012}. These models result in different BH seed masses; the remnants of Population III stars would create seeds with $M_{\rm BH}\sim100~M_\odot$, while stellar collisions and direct collapse would create BHs with $M_{\rm BH}\sim10^{3}\mbox{--}10^{5}~M_\odot$. 

While the early BH seeds at high redshift are too faint to be detected with current facilities \citep[e.g.][]{Volonteri:2016,Vito:2018,Schleicher:2018}, lower-mass galaxies, especially nearby dwarf galaxies, that harbor massive BHs can constrain BH seed models \cite[see][for reviews]{Greene:2020,Reines:2022}. The relatively quiet merger history of dwarf galaxies \citep{Bellovary:2011} as well as supernova feedback that may stunt BH growth \citep{Habouzit:2017,Angl:2017} can leave their BH masses close to their initial seed mass. 
Finding and studying BHs in dwarf galaxies is also important for understanding the role of both negative \citep{Manzano:2019} and positive \citep{Schutte:2022} AGN feedback in the low-mass regime.

There are multiple ways to search for BHs in the form of active galactic nuclei \citep[AGNs; see][for a review]{Ho:2008,Kewley:2019}. In the optical regime, narrow-line ratio diagnostic diagrams \citep[e.g.,][]{Baldwin:1981,Shirazi:2012} that differentiate between star forming (SF) and AGN ionizing spectral energy distributions (SEDs) have been employed  to identify AGN activity in lower-mass and dwarf galaxies \citep[e.g.,][]{Reines:2013,Moran:2014,Sartori:2015,Baldassare:2016}. Moreover, detection of broad H$\alpha$ emission \citep{Greene:2004,Greene:2007,Dong:2012,Reines:2013,Chilingarian:2018} can be indicative of the presence of dense gas in the broad line region (BLR) around a BH, thus suggestive of AGN activity in galaxies. High-ionization coronal emission lines, such as [\ion{Fe}{10}]$\lambda6374$ and [\ion{Ne}{5}]$\lambda3426$, can also be produced in the presence of massive BHs, thus an indicator of AGN activity \citep[e.g.,][]{penston:1984,prieto2002,satyapal:2008,goulding:2009,cerquiera:2021,Molina:2021,Molinafex:2021,Schmidt:1998ne5,Gilli:2010}.

There are selection biases associated with each AGN diagnostic, which results in the selection of different populations of galaxies. The narrow-line diagnostic diagrams typically observe high-accretion rate AGNs \citep{Greene:2020} and struggle with identifying low ionization nuclear emission regions (LINERs), low-luminosity AGNs (LLAGNs), and shock activity \citep{Ho:2008,Molina:2018,Kewley:2019}. This leaves a significant portion of lower-mass galaxies with lower accretion rates unexplored. Moreover, the radiation from the host galaxy can obscure AGN activity in SF galaxies \citep{Moran:2002,Groves:2006, Stasinska:2006,cann:2019}.
Thus, it is crucial to conduct searches for AGN activity that can minimize these effects and probe different populations of lower-mass galaxies. 

In this paper, we present a spectroscopic search for BH activity in low-mass galaxies utilizing data from the Galaxy And Mass Assembly (GAMA) survey Data Release 4 \citep[DR4;][]{Liske:2015,Driver:2022}. We analyze the spectra and search for AGN signatures in galaxies with stellar masses $M_\star\leq10^{10}M_\odot$ and redshifts $z\leq0.3$. Given that the GAMA spectroscopic survey covers different sky regions and is approximately two magnitudes deeper than the Sloan Digital Sky Survey (SDSS) spectroscopic survey \citep{york:2000}, where most previous optical searches have been conducted \citep[e.g.,][]{Greene:2007,Reines:2013,Moran:2014},
we aim to find novel AGN candidates in this stellar mass range.

We proceed by employing four AGN diagnostics, including 
two narrow-line diagnostic diagrams ([\ion{O}{3}]/H$\beta$ vs.\ [\ion{N}{2}]/H$\alpha$ and \ion{He}{2}/H$\beta$ vs.\ [\ion{N}{2}]/H$\alpha$), as well as searching for the [\ion{Fe}{10}]$\lambda6374$ and [\ion{Ne}{5}]$\lambda3426$ high-ionization coronal emission lines. This multiple diagnostic approach allows us to perform a more comprehensive search for AGN activity in this low-mass range, and thus potentially identify massive BHs from different populations of galaxies (e.g. in terms of their masses and colors).
We explain the data and our sample selection process in section \ref{sec:data} and the analysis of the GAMA spectra in section \ref{sec:analysis}. The results of each emission-line diagnostic and the host galaxy properties are included in sections \ref{sec:results} and \ref{sec:host_properties}, respectively. A summary and conclusions are presented in 
section \ref{sec:discussion_summary}. Here we assume a $\Lambda$CDM cosmology with $\Omega_m=0.3$, $\Omega_\Lambda=0.7$ and $H_0 = 70$ km s$^{-1}$ Mpc$^{-1}$.


\section{Data and Parent Sample of Low-Mass Galaxies}\label{sec:data}

\subsection{The GAMA Survey}\label{sec:GAMA_survey}
The GAMA Survey includes optical spectroscopy taken with the AAOmega multi-object spectrograph on the 3.9 m  Angelo-Australian Telescope \citep[AAT;][]{Saunders:2004,Smith:2004,Sharp:2006}. The spectrograph is equipped with a dual-beam setup that covers the wavelength range of 3730--8850 \AA\ with a dichroic split at 5700 \AA. The spectral resolution of the blue and red arms are 3.5 and 5.3 \AA, respectively, and the spectroscopic fibers are 2\arcsec\ in diameter. 
In this work, we utilize spectra and stellar masses released in GAMA DR4 covering three equatorial 60 deg$^2$ regions (G09, G12 and G15) and two southern $\sim50$ deg$^2$ regions (G02 and G23).
The combined limiting magnitude for the main survey objects in the equatorial and G23 regions is $r<19.65$ mag and the G02 region has a limiting magnitude of $r<19.8$ mag \citep{Baldry:2018,Driver:2022}.


\subsection{Parent Sample}\label{sec:sample_selection}
The GAMA database is stored in tables organized into data management units (DMUs)\footnote{\url{http://www.gama-survey.org/dr4/schema/}}.
The current GAMA spectra are provided in the \texttt{AATSpecAll v27} table in the \texttt{SpecCat} DMU \citep{Liske:2015}. We only use spectra with redshift estimates that are correct by the probability of at least 95\%. Additionally, if multiple GAMA spectra are matched to a single GAMA object, we use the spectrum that provides the best redshift for that object. We also exclude problematic spectra such as those that are affected by fringing and bad splicing.

To select our parent sample of low-mass galaxies, we impose a stellar mass cut of $M_\star \leq 10^{10} M_\odot$ using galaxy stellar masses provided by GAMA,
which are stored in the \texttt{StellarMasses} DMU \citep{Taylor:2011}. Stellar masses are obtained from stellar population fits to multiband SEDs. We utilize the mass estimates in the \texttt{StellarMassesGKV v24} table \citep{Driver:2022} for the equatorial and G23 survey regions, which uses matched-segment photometry across all bands derived  
from the Kilo-Degree Survey \citep[KiDS;][]{Kuijken:2019} and the  Visible and Infrared Survey Telescope for Astronomy Kilo-degree  Infrared Galaxy Public Survey \citep[VIKING;][]{Edge:2013}. 
Stellar masses for galaxies in the G02 survey region are provided in the \texttt{StellarMassesG02CFHTLS v24} and \texttt{StellarMassesG02SDSS v24} tables, which are based on multi-band SED fitting to Canada-France-Hawaii Telescope Lensing  \citep[CFHTLenS;][]{Heymans:2012} and SDSS photometry, respectively. We utilize the mass estimates given in the \texttt{StellarMassesG02CFHTLS v24} table, but use the \texttt{StellarMassesG02SDSS v24} table to remove galaxies with masses that are different by at least 0.3 dex in both tables. Finally, we apply the mass constraint of $10^{5} \leq M_* \leq 10^{10} M_\odot$, which results in 52,782 objects.

In addition to the stellar mass constraint, we also employ signal-to-noise (S/N) cuts using emission line measurements provided by GAMA.  In particular, we use the Gaussian-fit, emission-line fluxes and equivalent widths (EWs) from the \texttt{GaussFitSimple v05} table from the \texttt{SpecLineSFR} DMU \citep{Gordon:2017}. Following the \citet{Reines:2013} methodology, we impose the following requirements: the H$\alpha$, [\ion{O}{3}]~$\lambda5007$ and [\ion{N}{2}]~$\lambda6583$ lines must have ${\rm S/N}\geq3$ and ${\rm EW}>1$~\AA, and H$\beta$ must have ${\rm S/N}\geq2$. We also only include the objects with redshifts $z\leq0.3$ to ensure the [\ion{S}{2}] doublet is in the observed wavelength range. This leaves us with a parent sample consisting of 23,460 galaxies.


\section{Analysis of the GAMA Spectra}\label{sec:analysis}

In this work, we use 
a variety of optical emission line diagnostics
to search for AGN activity in our parent sample of low-mass emission-line galaxies. While we use the GAMA flux measurements to help define our parent sample, we create custom code to carry out our spectral analysis and search for AGN signatures. This includes fitting and subtracting the stellar continuum, separating broad and narrow H$\alpha$ and H$\beta$ components, and fitting various other emission lines. All of our custom code is written in the \texttt{Python} programming language\footnote{\url{https://www.python.org/}}. 
 

\subsection{Stellar Continuum Subtraction} \label{sec:continuum}
The stellar continuum, which significantly contributes to the observed spectra of the galaxies in our parent sample, needs to be removed before we can search for emission-line signatures of AGNs. Stellar light will generally contain absorption features and it is especially important to model and remove Balmer absorption lines when searching for broad H$\alpha$ or H$\beta$ emission that could signify dense gas orbiting a massive BH.

\begin{figure*}[htbp]
\centering
\includegraphics[width=\textwidth]{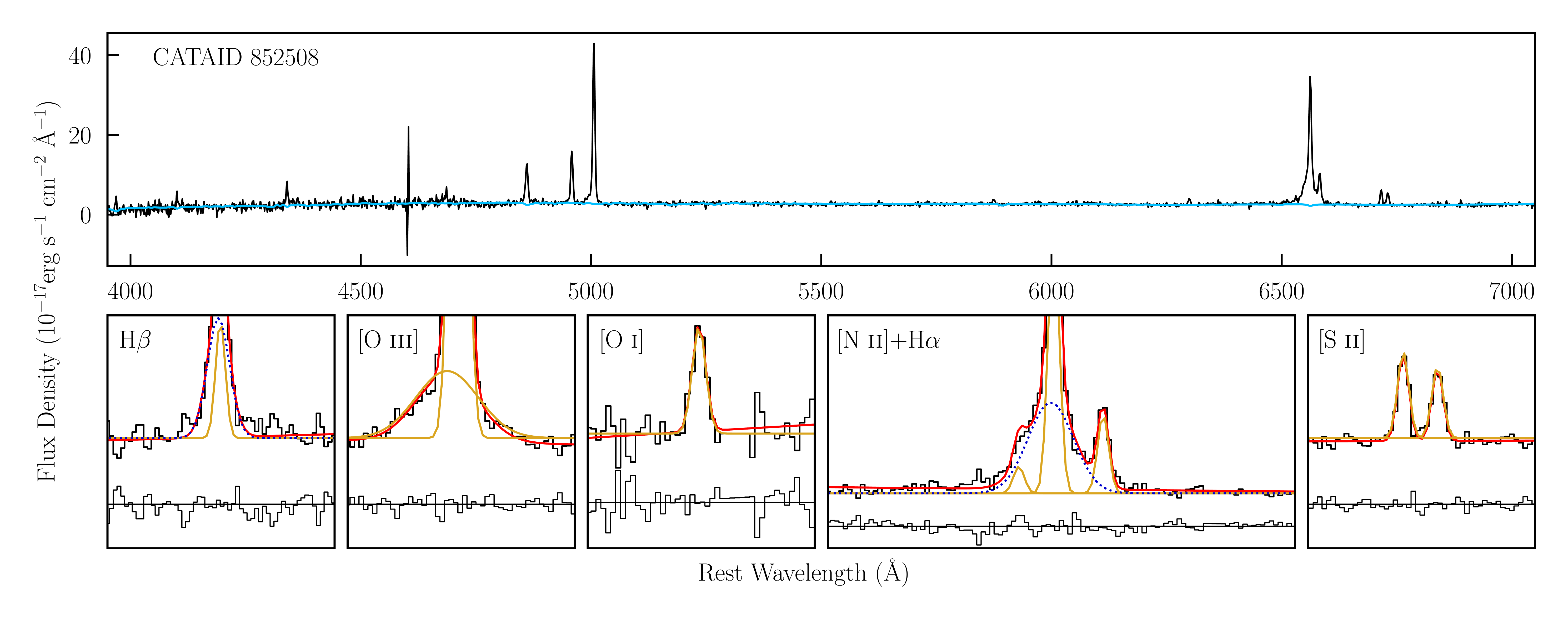}
\caption{An example of a redshift-corrected spectrum (top panel) for a broad-line AGN in our sample with chunks showing various emission-lines (bottom panels). This galaxy has CATAID 852508 and is among the [\ion{N}{2}]/H$\alpha$-selected AGN candidates that also has broad H$\alpha$ emission. In the top panel, we show the observed spectrum in black and the best-fit continuum + absorption line model in blue. In the bottom panels, we plot chunks of the emission-line spectra in black, the best fitting total model (Gaussian emission-line model + local linear continuum model) in red, and the individual Gaussian components in gold. The Gaussian broad H$\alpha$ and H$\beta$ components are plotted in dotted dark blue.} 
\label{fig:spec_line_sample}
\end{figure*}

We use the publicly available package \texttt{pPXF} \citep{Cappellari:2017} to find the best fit stellar continuum model for each spectrum. 
We use the \citet{Bruzual:2003} SSP models in the wavelength range of 3350 to 8850 \AA\ with spectral resolution of 3 \AA, which are calculated for 3 different metallicities ($Z=$ 0.008, 0.02, 0.05) and 10 different ages ($t = $ 0.005, 0.025, 0.1, 0.29, 0.64, 0.9, 1.4, 2.5, 5, and 11 Gyr). We model each spectrum with a combination of single-metallicity SSP models, modified by a low-order multiplicative polynomial to account for reddening by dust. This method yields acceptable continuum models as well as plausible velocity dispersions by \texttt{pPXF} for the majority of the objects in our sample. However, if the velocity dispersion is unrealistically large (200--1000 km/s), we refit the continuum including additive polynomials, which can change absorption line strengths and thereby help minimize template mismatch \citep{Cappellari:2017}. This was the case for 95 objects. We select the model metallicity with the smallest $\chi^2$ value. The majority of the galaxies in our sample ($\sim$ 72 \%) are best fitted by the sub-solar metallicity model $(Z=0.008)$. This is consistent with previous studies that show low-mass galaxies generally have low metallicities \citep[e.g.,][]{Tremonti:2004}. Since our primary goal is to measure the emission lines, we attempt good fits to the stellar continua but do not fully explore the parameter space. An example of a fitted galaxy spectrum is shown in the top panel of Figure \ref{fig:spec_line_sample}. In the end, we subtract the best-fit model from the data to achieve a pure emission-line spectrum.

\subsection{Emission Line Measurements} \label{sec:lines}
We use the \texttt{LMFIT} package in python \citep{lmfit} to model the emission lines with Gaussians. For each spectral region that we fit, we also include a linear component in the model to account for uncertainties associated with the initial stellar continuum fit. Examples of fitted emission lines are shown in the bottom panel of Figure \ref{fig:spec_line_sample}.

Following the methodology in \citet{Reines:2013} and references therein, we first fit the [\ion{S}{2}]$\lambda\lambda$6716,6731 doublet with single Gaussian models for each line in the doublet. We assume equal widths for the lines (in velocity space) and hold their relative laboratory wavelengths fixed. We then fit each line in the [\ion{S}{2}] doublet with a two-component Gaussian model. In this case, we additionally constrain the relative heights, widths and positions of the two components to be the same for both lines. We adopt the two-component Gaussian model if the reduced $\chi^2$ is at least 20\% lower than that of the single Gaussian model. Only 15 galaxies meet this criterion and require a two-component Gaussian model for the narrow line profile.

We then fit the [\ion{N}{2}]$\lambda\lambda$6548,6583 doublet and narrow H$\alpha$ line based on the parameters from the [\ion{S}{2}] emission-line model, as the [\ion{N}{2}] and narrow H$\alpha$ line profiles are well-matched to the [\ion{S}{2}] lines \citep{Filippenko:1988,Filippenko:1989,Ho:1997,Greene:2004}. The relative separation between the [\ion{N}{2}] lines is held fixed using their laboratory wavelengths and the flux ratio of [\ion{N}{2}]$\lambda$6583/[\ion{N}{2}]$\lambda6548$ is set to the theoretical value of 2.96. We fix the width of the lines in the [\ion{N}{2}] doublet (in velocity space) to that of the [\ion{S}{2}] lines, but let the width of the narrow H$\alpha$ line increase by as much as 25\%.  We scale the two-component [\ion{S}{2}] parameters for the 15 galaxies with two-component [\ion{S}{2}] models to fit the narrow-line emission of the [\ion{N}{2}] and H$\alpha$ group.
The [\ion{N}{2}]+H$\alpha$ complex is then fitted a second time with an additional broad H$\alpha$ component. If the computed reduced $\chi^2$ value is at least 20\% less than that of the narrow-line model, and the full-width at half maximum (FWHM) of the broad H$\alpha$ line is at least 500~km~s$^{-1}$ after correcting for the fiber-dependant instrumental resolution, we select the model with broad H$\alpha$ component. We fit the H$\beta$ line using the same method as the H$\alpha$ line.

We also model the [\ion{O}{3}]$\lambda$5007 and [\ion{O}{1}]$\lambda$6300 emission lines. Since the [\ion{O}{3}] line normally shows a broad, blue shoulder \citep[e.g.][]{Heckman:1981,Whittle:1985} and does not match the other line profiles \citep{Greene:2005}, we use an independent Gaussian model for the fitting process. We also need independent [\ion{O}{1}] model to accurately model the [\ion{Fe}{10}] line, which is discussed below.  We fit the [\ion{O}{3}] and [\ion{O}{1}] lines with one- and two-Gaussian models, and accept the two-component model if the measured reduced $\chi^2$ is lowered by at least 20\%.

We follow the methodology described in \citet{Molinafex:2021} to fit the [\ion{Fe}{10}]$\lambda$6374 line, which allows us to detect [\ion{Fe}{10}] even if it is blended with the [\ion{O}{1}]$\lambda$6363 line. We use the model parameters of the fitted [\ion{O}{1}]$\lambda$6300 line to describe the [\ion{O}{1}]$\lambda$6363 line.  
Specifically, we shift the model using the laboratory line wavelengths, assume the same width in velocity, and keep the flux ratios of [\ion{O}{1}]$\lambda$6300/[\ion{O}{1}]$\lambda$6363 $=$ 3. We also add a linear fit to the continuum in this spectral region.
Finally, we subtract the [\ion{O}{1}]$\lambda$6363 Gaussian component and the linear fit so we are left only with a potential 
[\ion{Fe}{10}] line, which we fit with a single Gaussian model.

We also search for \ion{He}{2} $\lambda$4686 and [\ion{Ne}{5}]$\lambda3426$ lines and fit a single Gaussian model to each line. Given the observed wavelength range of the GAMA survey, we only search for [\ion{Ne}{5}] emission in galaxies with redshift $z\geq0.15$.

We use the parameters from the Gaussian models to calculate the emission-line fluxes. 
We consider a line detected if the flux has a S/N $\geq3$.
In addition to the flux requirement, we require the line peak to be at least 3$\sigma$ above the noise for the relatively weak \ion{He}{2}$\lambda4686$, [\ion{O}{1}]$\lambda6300$, [\ion{Fe}{10}]$\lambda6374$, and [\ion{Ne}{5}]$\lambda3426$ lines, where the noise is determined as the root mean square (rms) of the continuum windows around the lines.
Finally, we visually inspect the AGN candidates that are flagged by our automated code and remove those that have spectra with missing pixel values within the emission lines, those affected by bad splicing or fringing, and those with bad fits to emission lines (e.g., noise or broad fits to the continuum). 
Given that the [\ion{N}{2}] and H$\alpha$ lines are fitted based on the [\ion{S}{2}] model parameters, a good fit the to the [\ion{S}{2}] lines is needed. However, if the flagged AGN candidates have strong [\ion{N}{2}] and H$\alpha$ detections, despite unreliable [\ion{S}{2}] detection, we keep them as potential AGN candidates in our final sample.


\begin{figure*}[tbph]
\centering
\includegraphics[width=0.8\textwidth]{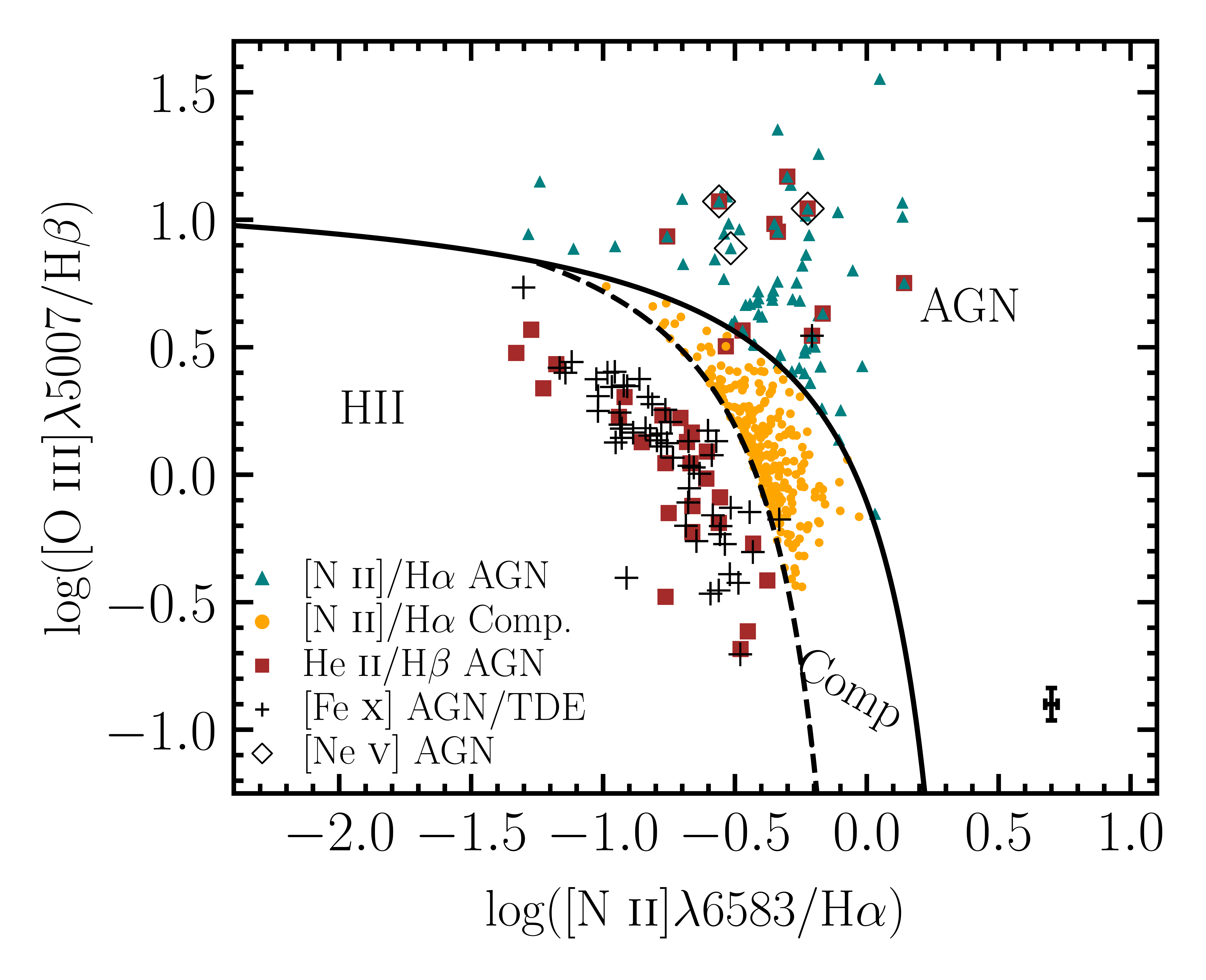}
\caption{The [\ion{O}{3}]/H$\beta$ vs.\ [\ion{N}{2}]/H$\alpha$ diagnostic diagram \citep{Baldwin:1981} for all 388 AGN candidates found in this work. The [\ion{N}{2}]/H$\alpha$ AGNs and Composites are shown as teal triangles and orange circles, respectively, while the \ion{He}{2}/H$\beta$ AGNs are shown as brown squares and [\ion{Fe}{10}]-selected AGNs as black pluses. The [\ion{Ne}{5}]-AGNs are added as unfilled back diamonds. Eleven of the \ion{He}{2}-selected AGNs, 2 of the [\ion{Fe}{10}]-selected AGNs, and all 3 of the [\ion{Ne}{5}]-selected AGNs overlap with the [\ion{N}{2}]/H$\alpha$ AGNs/Composites.}
\label{fig:AGNsamp}
\end{figure*}

\begin{deluxetable*}{ccDcccccccc}
\tabletypesize{\scriptsize}
\tablecaption{AGNs in Low-mass Galaxies}
\tablehead{
\colhead{CATAID}&\colhead{RA}& \multicolumn2c{DEC}  & \colhead{z} &\colhead{log ($M_*/M_\odot$)} & \colhead{$M_g$} & \colhead{$g-r$} &\colhead{[\ion{N}{2}]/H$\alpha$}&\colhead{\ion{He}{2}/H$\beta$}&\colhead{[\ion{Fe}{10}]$\lambda3426$}&\colhead{[\ion{Ne}{5}]$\lambda6364$}\\
\colhead{} & \colhead{} & \multicolumn2c{}& \colhead{} & \colhead{}  & \colhead{}  &\colhead{}&\colhead{location}&\colhead{location}&\colhead{Detection}&\colhead{Detection}
}
\decimalcolnumbers
\startdata
2273067 & 30.26775 & -4.17022 & 0.18946 &  9.95 & -20.20 & 0.42 &   Comp. &  \ldots &   \ldots &  \ldots \\
1460491 & 30.47504 & -7.00111 & 0.04907 &  9.43 & -18.50 & 0.47 &   Comp. &  \ldots &   \ldots &  \ldots \\
1459023 & 30.57854 & -7.08115 & 0.08646 &  9.60 & -18.89 & 0.49 &     AGN &  \ldots &   \ldots &  \ldots \\
1555169 & 30.60408 & -4.99215 & 0.19842 &  9.95 & -20.34 & 0.42 &     AGN &  \ldots &   \ldots &  \ldots \\
2258819 & 30.87408 & -9.49147 & 0.25170 &  9.60 & -20.66 & 0.27 &     AGN &  \ldots &   \ldots &  \ldots \\
1379310 & 31.24338 & -8.49599 & 0.11249 &  9.96 & -20.44 & 0.40 &   Comp. &  \ldots &   \ldots &  \ldots \\
1366459 & 31.42363 & -7.52362 & 0.13715 &  9.49 & -19.12 & 0.39 &      SF &  \ldots & Detected &  \ldots \\
1431173 & 31.50662 & -5.06840 & 0.29580 &  9.71 & -20.70 & 0.24 &     AGN &  \ldots &   \ldots &  \ldots \\
2229220 & 31.53125 & -5.62439 & 0.24479 &  9.90 & -20.58 & 0.31 &      SF &     AGN &   \ldots &  \ldots \\
1434726 & 31.77417 & -4.82812 & 0.28103 &  9.99 & -20.88 & 0.33 &   Comp. &  \ldots &   \ldots &  \ldots \\
\enddata
\tablecomments{Galaxy properties for the low-mass AGNs. The values given in columns 1-7 are obtained from GAMA DR4 and assume $h = 0.7$. Column 1: Unique ID of the GAMA object. Columns 2--3: The right ascension and declination (in degrees) of the spectrum (J2000). Column 4: Redshift. Columns 5--7: The log galaxy stellar mass in units of $M_\odot$, absolute $g-$band magnitude, and $g-r$ color. All values are obtained from the \texttt{StellarMassesG02CFHTLS v24} and \texttt{StellarMassesGKV v24} tables \citep{Taylor:2011,Bellstedt:2020}. Columns 8--9: Our classifications of the galaxy in each of the narrow-line diagnostic diagrams. Columns 10--11: The [\ion{Fe}{10}] and [\ion{Ne}{5}] coronal line detections in this work. 
A three-dot ellipsis indicates that no line is detected.
The entirety of Table~\ref{tab:gal_prop} is published in the electronic edition of {\it The Astrophysical Journal}. We show a portion here to give information on its form and content.
}
\label{tab:gal_prop}
\end{deluxetable*}

\begin{deluxetable*}{ccccccccccccc}
\tabletypesize{\scriptsize}
\setlength{\tabcolsep}{1.5pt}
\renewcommand{\arraystretch}{1.}
\tablewidth{0pt}
\tablecaption{Emission-line Fluxes}
\tablehead{
\colhead{CATAID}&\colhead{\ion{Ne}{5}$\lambda3426$}&\colhead{\ion{He}{2}$\lambda4686$} &\colhead{(H$\beta$)$_n$}&\colhead{(H$\beta$)$_b$}& \colhead{[\ion{O}{3}]$\lambda5007$}& \colhead{[\ion{O}{1}]$\lambda6300$} &\colhead{[\ion{Fe}{10}]$\lambda6374$}& \colhead{(H$\alpha$)$_n$}  &\colhead{(H$\alpha$)$_b$}&\colhead{[\ion{N}{2}]$\lambda6583$} & \colhead{[\ion{S}{2}]$\lambda6716$} & \colhead{[\ion{S}{2}]$\lambda6731$}}
\decimalcolnumbers
\startdata
2273067 & \ldots & \ldots &  41(6) & \ldots &  160(3) &  18(2) & \ldots &  214(6) & \ldots &   37(4) &   43(3) &  37(3) \\
1460491 & \ldots & \ldots & 85(25) & \ldots & 142(23) & \ldots & \ldots & 325(16) & \ldots & 117(13) & 116(17) & 96(17) \\
1459023 & \ldots & \ldots &  87(7) & \ldots &  156(7) & \ldots & \ldots &  303(4) & \ldots &  241(3) &  105(5) &  55(5) \\
1555169 & \ldots & \ldots &   6(7) & 71(10) &   27(3) & \ldots & \ldots &  196(4) & \ldots &   68(3) &   38(2) &  23(2) \\
2258819 & \ldots & \ldots &   4(1) &  42(3) &   53(2) & \ldots & \ldots &   32(2) & 196(4) &   17(1) &  \ldots & \ldots \\
1379310 & \ldots & \ldots & 272(6) & \ldots &  238(8) &  53(3) & \ldots & 1356(8) & \ldots &  616(6) &  223(4) & 182(4) \\
1366459 & \ldots & \ldots &  33(3) & \ldots &   18(2) & \ldots &  24(3) &  110(3) & \ldots &   25(2) &   30(3) &  10(3) \\
1431173 & \ldots & \ldots &  27(2) & \ldots &  209(3) & \ldots & \ldots &  186(3) & \ldots &   21(2) &   26(4) &  22(4) \\
2229220 & \ldots &  38(4) &  32(3) & \ldots &   26(3) & \ldots & \ldots &  119(3) & \ldots &   33(3) &   17(4) &  19(4) \\
1434726 & \ldots & \ldots &  36(4) & \ldots &   54(3) & \ldots & \ldots &  173(5) & \ldots &   73(4) &   32(5) &  26(5) \\
\enddata
\tablecomments{Measured emission line fluxes for our sample of low-mass AGNs. Column 1: Unique ID of the GAMA object. Columns 2--13: The emission-line fluxes in units of 10$^{-17}$ erg s$^{-1}$ cm$^{-2}$ with the errors shown in parentheses.
No extinction correction has been applied. The subscripts n and b refer to the narrow and broad components, respectively. We do not show the flux values of the weaker lines [\ion{N}{2}] and [\ion{O}{1}] doublets, since their fluxes are fixed to be weaker by factors of 2.96 and 3, respectively. A three-dot ellipsis indicates that no line is detected.
The entirety of Table \ref{tab:flux} is published in the electronic edition of {\it The Astrophysical Journal}. We show a portion here to give information on its form and content. }
\label{tab:flux}
\end{deluxetable*}

\section{AGN Selection}
\label{sec:results}

In this work, we search for various optical spectroscopic indicators of AGN activity using the emission line measurements described above. 
In order to provide a comprehensive search for AGN activity, we employ four different AGN diagnostics that we consider to be relatively robust in the low-mass regime.  These include the [\ion{O}{3}]/H$\beta$ vs.\ [\ion{N}{2}]/H$\alpha$ and \ion{He}{2}/H$\beta$ vs.\ [\ion{N}{2}]/H$\alpha$ 2D narrow emission line ratio diagrams 
\citep{Baldwin:1981,Shirazi:2012}, as well as searching for [\ion{Fe}{10}] and [\ion{Ne}{5}] and coronal-line emission \citep[][]{Molinafex:2021,Schmidt:1998ne5,Gilli:2010}. 
We also search for broad H$\alpha$ emission \citep[e.g.,][]{Greene:2005,Reines:2013,Chilingarian:2018} in our parent sample, but only include
the broad-line AGN candidates that overlap with other AGN diagnostics in this work since broad H$\alpha$ can also be produced by transient stellar phenomena (e.g., Type II supernovae, in star-forming galaxies with low masses; \citealt{Baldassare:2016}). 
We describe each of the four AGN diagnostics below, and present the results of applying each diagnostic to our parent sample of low-mass emission-line galaxies (also see Figure \ref{fig:AGNsamp}). The galaxy properties of the AGN candidates and their respective emission-line flux measurements are listed in Tables \ref{tab:gal_prop} and \ref{tab:flux}, respectively.


\subsection{[\texorpdfstring{\ion{O}{3}}{TEXT}]/H\texorpdfstring{$\beta$}{TEXT} vs. [\texorpdfstring{\ion{N}{2}}{TEXT}]/H\texorpdfstring{$\alpha$}{TEXT}}\label{sec:nii}

The photoionizing continuum from an AGN contains a larger fraction of high-energy photons relative to hot stars, which results in extended partially ionized regions in AGNs. In these regions, lines such as [\ion{N}{2}]$\lambda$6583, [\ion{S}{2}]$\lambda\lambda$6716,6731, and [\ion{O}{1}]$\lambda$6300 are produced by collisional excitation. This results in larger intensities of these lines with respect to H$\alpha$ in the narrow-line emission from AGNs than in \ion{H}{2} regions, which allows them to be separated in emission-line diagnostic diagrams.

\begin{figure*}[ptbh]
\centering
\includegraphics[width=\textwidth]{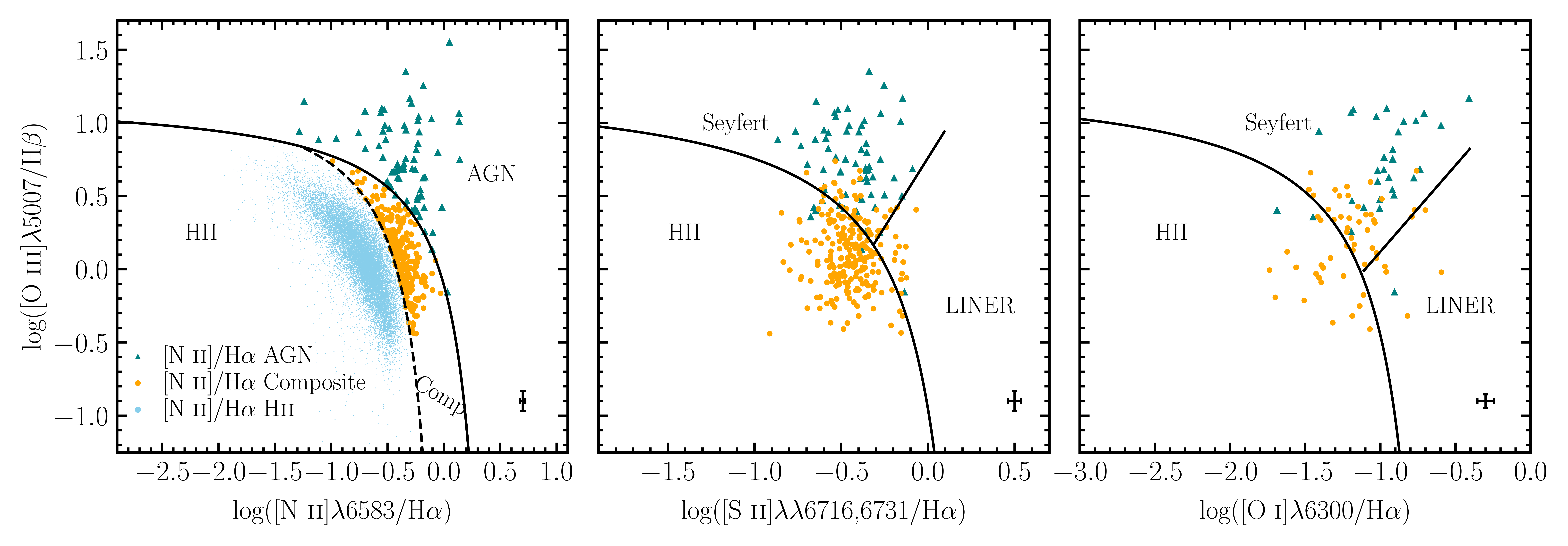}
\caption{The [\ion{O}{3}]/H$\beta$ vs.\ [\ion{N}{2}]/H$\alpha$ narrow-line diagnostic diagram (left panel) for galaxies with stellar masses $M_\star\leq10^{10} M_\odot$ in the GAMA survey using the classification scheme summarized in \citet{Kewley:2006}. We show the 71 AGN and 238 composite galaxies as teal triangles and orange circles, respectively, and those consistent with star-forming galaxies as sky blue points. The middle and right panels show the AGN and composite objects in the [\ion{O}{3}]/H$\beta$ vs. [\ion{S}{2}]/H$\alpha$ and [\ion{O}{1}]/H$\alpha$ diagrams. Only AGN candidates with reliable  [\ion{S}{2}] and/or [\ion{O}{1}] detection are plotted in the middle and right panels. Characteristic error bars are located in the lower right region of each panel. See Section~\ref{sec:nii} for details. 
}
\label{fig:bpt_nii}
\end{figure*}

The [\ion{O}{3}]/H$\beta$ vs.\ [\ion{N}{2}]/H$\alpha$ diagnostic diagram \citep{Baldwin:1981} has been widely used to separate SF galaxies from AGN-dominated ones. This diagram is metallicity sensitive, with SF galaxies varying in abundance from low metallicity (low [\ion{N}{2}]/H$\alpha$ ratio, high [\ion{O}{3}]/H$\beta$ ratio) to high metallicity (high [\ion{N}{2}]/H$\alpha$ ratio, low [\ion{O}{3}]/H$\beta$ ratio), while shocks and AGN-dominated galaxies generally have higher ratios of [\ion{O}{3}]/H$\beta$ and [\ion{N}{2}]/H$\alpha$. This results in a clear separation between SF galaxies and those with an AGN contribution in the general population of galaxies  \citep[e.g., ][]{Kewley:2019}. However, this diagnostic diagram can struggle with identifying AGNs in low-mass galaxies, which tend to have lower metallicities than more massive ones. In other words, low-metallicity AGNs overlap with low-metallicity starbursts in this diagram \citep{Groves:2006} and so these AGNs may be missed. Nevertheless, this diagram appears to be robust at identifying bona-fide AGNs in the low-mass regime \citep{Reines:2013,Baldassare:2017}.

We employ this diagram as our first AGN indicator as shown in the left panel of Figure \ref{fig:bpt_nii}. We use the classification scheme outlined in \citet{Kewley:2006}, where star-forming/\ion{H}{2} galaxies fall below the empirical composite line from \citet{Kauffmann:2003}, AGN-dominated galaxies fall above the theoretical extreme starburst line from \citet{Kewley:2001}, and composite galaxies fall in between the two lines.  We identify 71 AGNs and 238 composite galaxies in our parent sample by using this diagram. 

We also plot these AGN and composite galaxies in the [\ion{O}{3}]/H$\beta$ vs.\ [\ion{S}{2}]/H$\alpha$ and [\ion{O}{1}]/H$\alpha$ diagrams \citep{Veilleux:1987} as shown in the middle and right panels of Figure \ref{fig:bpt_nii}. In these diagrams, we use the classification scheme in \citet{Kewley:2006} where the star-forming galaxies and the AGN candidates are separated by the theoretical extreme starburst line from \citet{Kewley:2001} and the Seyfert-like and LINER-like galaxies by the Seyfert-LINER line. We find that 298/309 and 89/309 of the AGNs and composites have reliable [\ion{S}{2}] and [\ion{O}{1}] detections (see section \ref{sec:lines}), out of which 39\%  fall in the AGN region of the [\ion{S}{2}]/H$\alpha$ diagram and 74\% are AGN-like in the [\ion{O}{1}]/H$\alpha$ diagram. There are also 47 objects that show AGN activity in all three diagrams. Moreover, some of the AGNs/Composites selected by this diagnostic have additional AGN indicators (see Figure \ref{fig:AGNsamp} and the following subsections).


\subsection{\texorpdfstring{\ion{He}{2}/H$\beta$}{TEXT} vs. [\texorpdfstring{\ion{N}{2}]/H$\alpha$}{TEXT}}
\label{sec:he2}

Nebular \ion{He}{2} emission has a relatively high ionization potential (54.4 eV) and therefore can also be produced by a hard ionizing spectrum, which may indicate AGN activity. The \ion{He}{2}/H$\beta$ vs. [\ion{N}{2}]/H$\alpha$ diagram proposed by \citet{Shirazi:2012} has been used to separate SF galaxies from AGN-dominated ones in dwarf galaxies \citep[][]{Sartori:2015}. While \ion{He}{2} emission can originate from AGN activity, stellar processes can also produce this line; thus care is needed when using this diagnostic. 

\begin{figure}[tbph]
\centering
\includegraphics[width=0.45\textwidth]{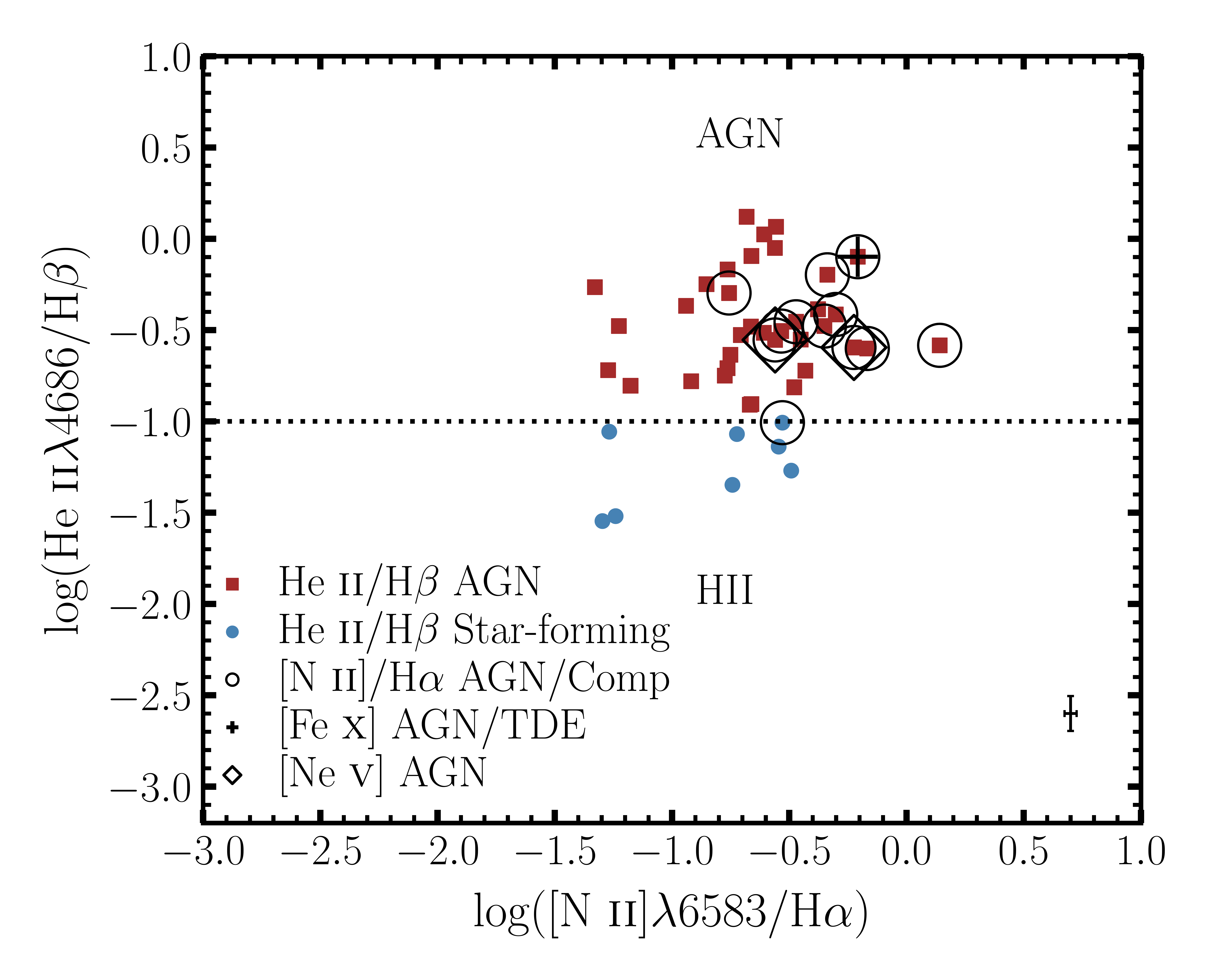}
\caption{The \ion{He}{2}/H$\beta$ vs. [\ion{N}{2}]/H$\alpha$ diagnostic diagram. We apply the criterion log(\ion{He}{2}/H$\beta$)$>-1$ (see section \ref{sec:he2}) to separate star-forming galaxies from galaxies with AGN activity, which is shown as the black dotted line. 36 out of the 44 galaxies with detectable \ion{He}{2} emission in our parent sample fall above this line, which we show as brown squares, while the SF galaxies are shown as steel blue circles. 12 of the galaxies in this sample are also AGNs/Composites in the [\ion{N}{2}]/H$\alpha$ diagnostic, which we indicate with unfilled circles, while 1 AGN is observed with [\ion{Fe}{10}] emission and 2 AGNs are also observed with [\ion{Ne}{5}] emission, which are plotted as a plus and unfilled black diamonds, respectively. The characteristic error bars are added in the lower right of this diagram.}
\label{fig:he2_diag}
\end{figure}

We search for \ion{He}{2} emission in our parent sample of low-mass galaxies and identify 44 galaxies with detected emission, out of which 12 overlap with the [\ion{N}{2}]/H$\alpha$-selected AGNs and composites. We select the \ion{He}{2}/H$\beta$ AGN candidates in our sample by employing the criterion proposed in \citet{Molinafex:2021}, log(\ion{He}{2}/H$\beta$)$>-1$, as shown in Figure \ref{fig:he2_diag}. This limit is expected to be higher than that produced by X-ray binaries (XRBs) or Wolf-Rayet (WR) stars \citep{Schaerer:2019} and is slightly stricter than the criteria presented in \citet{Shirazi:2012}. We find that 36 of the \ion{He}{2}-emitting galaxies meet this criterion, out of which 10 are also [\ion{N}{2}]/H$\alpha$ AGNs and 1 is a composite object. The remaining \ion{He}{2}-emitting galaxy among the [\ion{N}{2}]/H$\alpha$-selected composite galaxies, while strictly in the \ion{H}{2} part of the diagram, is consistent with a \ion{He}{2}-selected AGN within the measurement uncertainties (see Figure \ref{fig:he2_diag}). One of the \ion{He}{2}/H$\beta$ AGNs has [\ion{Fe}{10}] emission and 2 have [\ion{Ne}{5}] emission (see sections \ref{sec:fex} and \ref{sec:ne5}), all three of which are also [\ion{N}{2}]/H$\alpha$ AGNs. In appendix \ref{appendix:lines}, we show the observed spectra for a selection of these AGN candidates in Figure \ref{fig:heii_spectra}, and the \ion{He}{2} emission line fits for all 36 \ion{He}{2}/H$\beta$-selected AGNs in Figure \ref{fig:spec_he2}.

Given that the majority (25/36) of the \ion{He}{2}/H$\beta$ AGN candidates are SF in the [\ion{N}{2}]/H$\alpha$ diagram, we further investigate these systems. 
First, we visually search for WR features in the spectra \citep{Conti:1991,Schaerer:1999} such as the blue and red bumps that appear around 4650 \AA\ and 5808 \AA. We do not find WR signatures in these galaxies, and thus conclude that either WR stars are not responsible for the observed \ion{He}{2} emission or any potential WR signatures are not detectable in the GAMA spectra.

\begin{figure*}[!th]
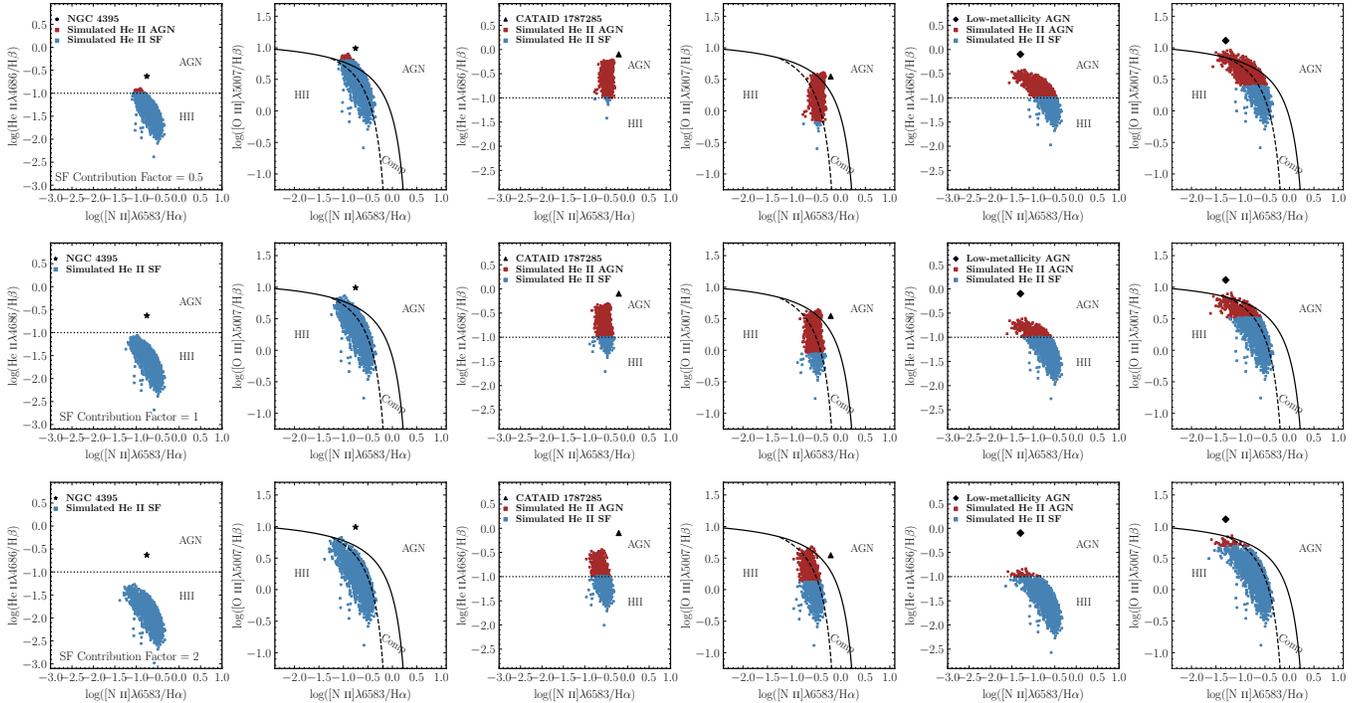

\centering
\includegraphics[width=\textwidth]{figures/he2_comparison_scale05.pdf}
\includegraphics[width=\textwidth]{figures/he2_comparison_scale1.pdf}
\includegraphics[width=\textwidth]{figures/he2_comparison_scale2.pdf}
\caption{Simulated emission line ratios in the \ion{He}{2}/H$\beta$ and [\ion{O}{3}]/H$\beta$ vs.\ [\ion{N}{2}]/H$\alpha$ diagrams from various combinations of AGN + SF line fluxes demonstrating that it is possible to have \ion{He}{2}-selected AGNs that fall in the SF region of the [\ion{N}{2}]/H$\alpha$ diagram. We show the results using three AGNs: NGC 4395 (first two columns), a \ion{He}{2}-selected AGN from this work (CATAID 1787285; middle two columns), and a mock low-metallicity AGN (last two columns). The SF line fluxes are scaled by the ratio of the [\ion{O}{3}]$\lambda5007$ line for each AGN to each SF galaxy by ``SF contribution factors" of 0.5 (first row), 1 (second row), and 2 (third row) before the addition.  
See Section \ref{sec:he2} for details. 
}
\label{fig:spec_he2_sample}
\end{figure*}

Next, we investigate whether it is possible to have a \ion{He}{2}/H$\beta$-selected AGN that is also SF in the [\ion{N}{2}]/H$\alpha$ diagram by combining a variety of AGN spectra with SF spectra. We begin by linearly adding emission-line fluxes from a well-known AGN in a dwarf galaxy, NGC 4395 \citep{Filippenko:1989,Filippenko:2003}, to $\sim 3000$ [\ion{N}{2}]/H$\alpha$ SF galaxies. We select these objects from our parent sample of low-mass galaxies described in section \ref{sec:sample_selection}, and require a S/N $>3$ for all the emission lines of interest (H$\beta$, [\ion{O}{3}], [\ion{N}{2}], H$\alpha$). None of these SF objects have detectable \ion{He}{2} emission, and each line of interest is scaled by the ratio of the [\ion{O}{3}]$\lambda5007$ line for NGC 4395 to that of each SF galaxy by factors of 0.5, 1, and 2 (i.e., a scale factor of 0.5 indicates a lower amount of star formation contribution to the synthesized line ratios). We then linearly add the scaled emission-line fluxes to those of NGC 4395 and plot the resulting emission-line ratios in the \ion{He}{2}/H$\beta$ and [\ion{O}{3}]/H$\beta$ vs.\ [\ion{N}{2}]/H$\alpha$ diagrams as shown in the first two columns of Figure \ref{fig:spec_he2_sample}. None of the constructed line ratios in this test are simultaneously \ion{He}{2}/H$\beta$ AGNs and [\ion{N}{2}]/H$\alpha$ SF galaxies. However, if the \ion{He}{2} line fluxes were stronger than that of NGC 4395 by at least a factor of 1.2, 1.4, and 2.1 for the scale factors of 0.5, 1, and 2, respectively, there would be galaxies that are both \ion{He}{2}/H$\beta$ AGNs and SF in the [\ion{N}{2}]/H$\alpha$ diagram.

In the next test, we employ the same methodology described above, but instead of using the emission-line fluxes from NGC 4395, we use 10 galaxies in our sample that are both [\ion{N}{2}]/H$\alpha$ and \ion{He}{2}/H$\beta$-selected AGNs. In 4/10 of these case studies, we find that there are objects that have emission-line ratios that are simultaneously AGN-like in the \ion{He}{2}/H$\beta$ diagram and SF in [\ion{N}{2}]/H$\alpha$ diagram. We show an example in the two middle columns of Figure \ref{fig:spec_he2_sample}. The AGN in this Figure (CATAID 1787285) has 
an \ion{He}{2}/[\ion{O}{3}] ratio $\sim 10$ times higher than NGC 4395. The fraction of objects that are \ion{He}{2}/H$\beta$ AGNs and SF in the [\ion{N}{2}]/H$\alpha$ diagram ranges from 15 \% to 50 \% depending on the SF contribution scale factor. However, we note that there is a continuum of objects reaching up into the composite region of the [\ion{N}{2}]/H$\alpha$ diagram, which is not seen in our sample of \ion{He}{2}-selected AGNs (Figure \ref{fig:AGNsamp}). The majority of the \ion{He}{2}-selected AGNs in our sample fall in the SF region of the [\ion{N}{2}]/H$\alpha$ diagram, with only 1 composite object and a handful of AGNs. Motivated by this, we next investigate the impact of using a low-metallicity AGN on the simulated line ratios.

We carry out the final test with a mock low-metallicity AGN, setting log([\ion{N}{2}]/H$\alpha)=-1.3$ and log([\ion{O}{3}]/H$\beta)=1.1$. We use the same value of log \ion{He}{2}/H$\beta = -0.1$ as the GAMA object in Figure \ref{fig:spec_he2_sample}. The simulated AGN+SF line ratios are shown in the last two columns of Figure \ref{fig:spec_he2_sample}.
In this case we again find \ion{He}{2}/H$\beta$ AGNs that are SF in the [\ion{N}{2}]/H$\alpha$ diagram for all the SF contribution scale factors.

The results given above indicate that factors such as metallicity, star formation versus AGN contribution, and the \ion{He}{2}/[\ion{O}{3}] ratio can impact where objects fall in the diagnostic diagrams. While this exercise has demonstrated that is certainly possible, and perhaps likely, that the detected \ion{He}{2} emission in this work is driven by AGN activity, follow-up observations would be useful to confirm these \ion{He}{2}-selected AGNs that are SF in the [\ion{N}{2}]/H$\alpha$ diagram.


\subsection{[\texorpdfstring{\ion{Fe}{10}}{TEXT}]\texorpdfstring{$\lambda$}{TEXT}6374 Coronal Line Emission}
\label{sec:fex}

The presence of [\ion{Fe}{10}]$\lambda$6374 emission with high ionization potential \citep[262.1 eV;][]{Oetken:1977} can be indicative of AGN activity in galaxies \citep[e.g.,][]{penston:1984,prieto2002,satyapal:2008,goulding:2009,cerquiera:2021}. Recent studies presented in \citet{Kimbro:2021} and \citet{Molina:2021} also confirmed the existence of
[\ion{Fe}{10}]$\lambda$6374 line from accreting BHs in dwarf galaxies. We note that, however, this line is usually weak and thus hard to detect.

We search for the [\ion{Fe}{10}]$\lambda$6374 line in our parent sample of galaxies and identify 56 reliable detections, out of which 1 overlaps with the [\ion{N}{2}]/H$\alpha$-selected AGNs and 1 is a composite object. Moreover, the [\ion{N}{2}]/H$\alpha$-selected AGN is also among the \ion{He}{2}/H$\beta$ AGNs.
We show the observed spectra for a selection of these objects in Figure \ref{fig:fex_spectra} and the [\ion{Fe}{10}] and [\ion{O}{1}] doublet emission-line fits for all 56 galaxies in Figure~\ref{fig:oi_fex} in appendix \ref{appendix:lines}.
The luminosity of the [\ion{Fe}{10}] lines in our sample span a range of $\sim 10^{38}-$10$^{41}$ erg s$^{-1}$, with a median of $10^{39.6}$ erg s$^{-1}$. Given these luminosities, there are two main sources that could explain the observed [\ion{Fe}{10}] emission: AGNs or tidal disruption events (TDEs), which is where a massive BH tidally disrupts a star. AGN activity can produce the [\ion{Fe}{10}] line as a result of gas photoionized by the AGN continuum \citep[e.g., ][]{Nussbaumer:1970,Pier:1995,Negus:2021}, or radiative shock waves emitted by radio jets from the AGN \citep[e.g., ][]{wilson:1999,Molina:2021}. A class of tidal disruption events (TDEs) called extreme coronal line emitters (ECLEs) also produce coronal-line emission with L$_{[\rm FeX]}\sim$ 10$^{38-40}$ erg s$^{-1}$ \citep{Komossa:2008,Wang:2011,Wang:2012}. 

Other potential origins of [\ion{Fe}{10}] emission are discussed in \citet{Molinafex:2021}, but these all fail to explain the high luminosities observed here. For example, supernovae rarely produce coronal lines and their luminosities are generally orders of magnitude lower than those observed in our sample. Even one of the most extreme examples, SN 2005ip, had a peak [\ion{Fe}{10}]$\lambda$6374 luminosity of just 2$\times 10^{37}$ erg s$^{-1}$ \citep{smith:2009}.
Therefore, we conclude that the observed [\ion{Fe}{10}] emission in our sample of low-mass galaxies is indicative of AGN activity or tidal disruption events (TDEs), both of which require the presence of a massive BH.

\begin{figure*}[!th]
\centering
\includegraphics[width=0.85\textwidth]{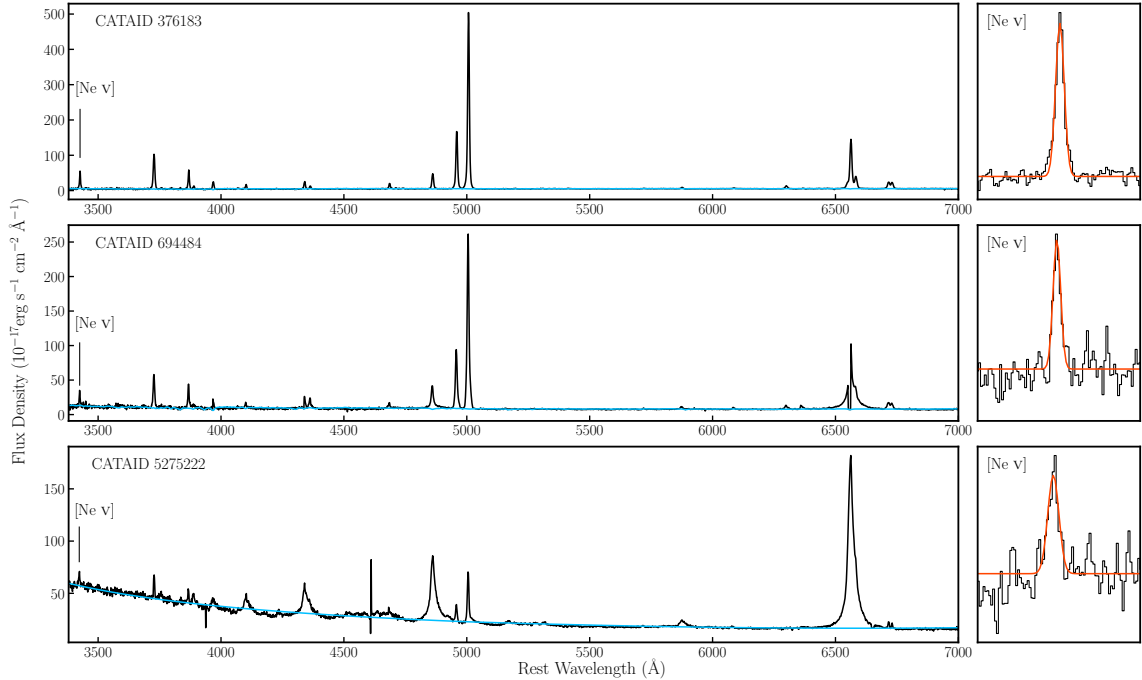}
\caption{Redshift-corrected spectra and chunks showing [\ion{Ne}{5}]$\lambda3426$ emission for the 3 strong [\ion{Ne}{5}]-emitting galaxies in this work. Here we plot the best-fitting continuum model and individual Gaussian component as sky blue and orange-red, respectively. These galaxies are also [\ion{N}{2}]/H$\alpha$ AGNs and 2 of them (CATAIDs 376183 and 694484) are \ion{He}{2}/H$\beta$ AGNs. 
See section \ref{sec:ne5} for details.}
\label{fig:spec_ne5}
\end{figure*}

\subsection{[\texorpdfstring{\ion{Ne}{5}}{TEXT}]\texorpdfstring{$\lambda$}{TEXT}3426 Coronal Line Emission}
\label{sec:ne5}
The presence of coronal lines with high ionization energies such as [\ion{Ne}{5}]$\lambda$3426 ($\sim97$ eV) are generally considered strong indicators of AGN activity \citep{Schmidt:1998ne5,Gilli:2010}. However, this line has also been found in star-forming galaxies \citep{Izotov:2004}, and it is generally weak and hard to detect.

We search for [\ion{Ne}{5}] emission in our parent sample of low-mass galaxies and identify 5 galaxies with such emission. However, we cut 2 of the objects with marginal [\ion{Ne}{5}] detections and spectra that do not show any other AGN signatures. The remaining 3 [\ion{Ne}{5}]-emitting galaxies are [\ion{N}{2}]/H$\alpha$ AGNs, 2 of which are also \ion{He}{2}/H$\beta$-selected AGN candidates. We show the observed spectra as well as the [\ion{Ne}{5}] emission-line fits for these galaxies in Figure \ref{fig:spec_ne5}.  The luminosities of the [\ion{Ne}{5}] lines are in the range of $10^{40.9-41.4}$ erg s$^{-1}$.


\subsection{Broad H\texorpdfstring{$\alpha$}{TEXT} Emission and Black Hole Masses}
\label{sec:broadha}

Dense gas orbiting in the vicinity of a BH can produce broad-line emission, such as broad H$\alpha$, and can be used to estimate the mass of the central BH \citep{Greene:2005}. However, in low-mass galaxies, broad H$\alpha$ emission from stellar-processes such as supernovae can mimic that of an AGN. Thus, transient broad H$\alpha$ emission that disappears over time likely indicates a supernova origin, whereas persistent broad H$\alpha$ favors an AGN origin
\citep[e.g.,][]{Baldassare:2016}.

\begin{figure*}[htbp]
\centering
\includegraphics[width=\textwidth]{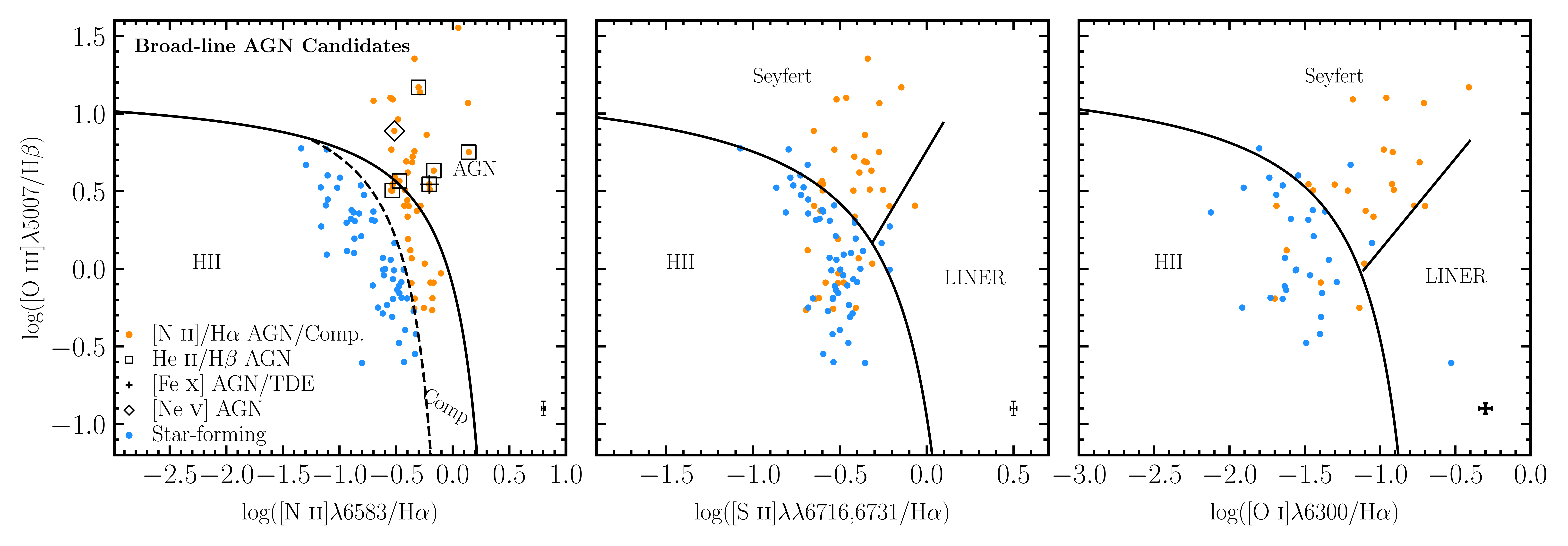}
\caption{Galaxies with detectable broad H$\alpha$ emission from our parent sample with stellar masses $M_\star\leq10^{10} M_\odot$ in the [\ion{O}{3}]/H$\beta$ vs.\ [\ion{N}{2}]/H$\alpha$ diagram (left panel), [\ion{O}{3}]/H$\beta$ vs.\ [\ion{S}{2}]/H$\alpha$ diagram (middle panel), and [\ion{O}{3}]/H$\beta$ vs.\ [\ion{O}{1}]/H$\alpha$ diagram (right panel). As in Figure \ref{fig:bpt_nii}, we use the classification scheme summarized in \citet{Kewley:2006}. We find 103 galaxies with broad H$\alpha$ emission, out of which 47 fall in the AGN/Composite regions of the [\ion{N}{2}]/H$\alpha$ diagram (orange points). Additionally, 6 of these are \ion{He}{2}-selected AGNs (black squares), 1 is
an [\ion{Fe}{10}]-selected AGN (black plus), and 1 is a [\ion{Ne}{5}]-selected AGN (black diamond). 
The broad H$\alpha$ objects falling the star-forming part of the [\ion{N}{2}]/H$\alpha$ diagram (blue points) do not overlap with any of the narrow-line diagnostics used in this work.  Only the objects with reliable [\ion{S}{2}] and/or [\ion{O}{1}] detections are plotted in the middle and right panels. Characteristic error bars are located in the lower right region of each panel. See section \ref{sec:broadha} for details. 
}
\label{fig:broad_bpt}
\end{figure*}

We search for broad H$\alpha$ emission in our parent sample of low-mass galaxies and identify 103 galaxies with such emission. As shown in Figure \ref{fig:broad_bpt}, 47 of these galaxies are in our [\ion{N}{2}]/H$\alpha$-selected AGN and composite sub-sample. Additionally, 7 of these 47 objects show additional AGN-like signatures: 6 are also \ion{He}{2}-selected AGNs, 1 has observed [\ion{Ne}{5}] emission, and 1 is both a \ion{He}{2}-selected AGN and has detectable [\ion{Fe}{10}] emission, while 1 of the 47 galaxies is SF in the \ion{He}{2}/H$\beta$ diagram. There is also one broad-line AGN candidate that is consistent with SF in the [\ion{N}{2}]/H$\alpha$ and \ion{He}{2}/H$\beta$ diagrams.
The remaining galaxies do not overlap with any of the diagnostics we employ in this work. 

\begin{figure*}[tbph]
\centering
\includegraphics[width=\textwidth]{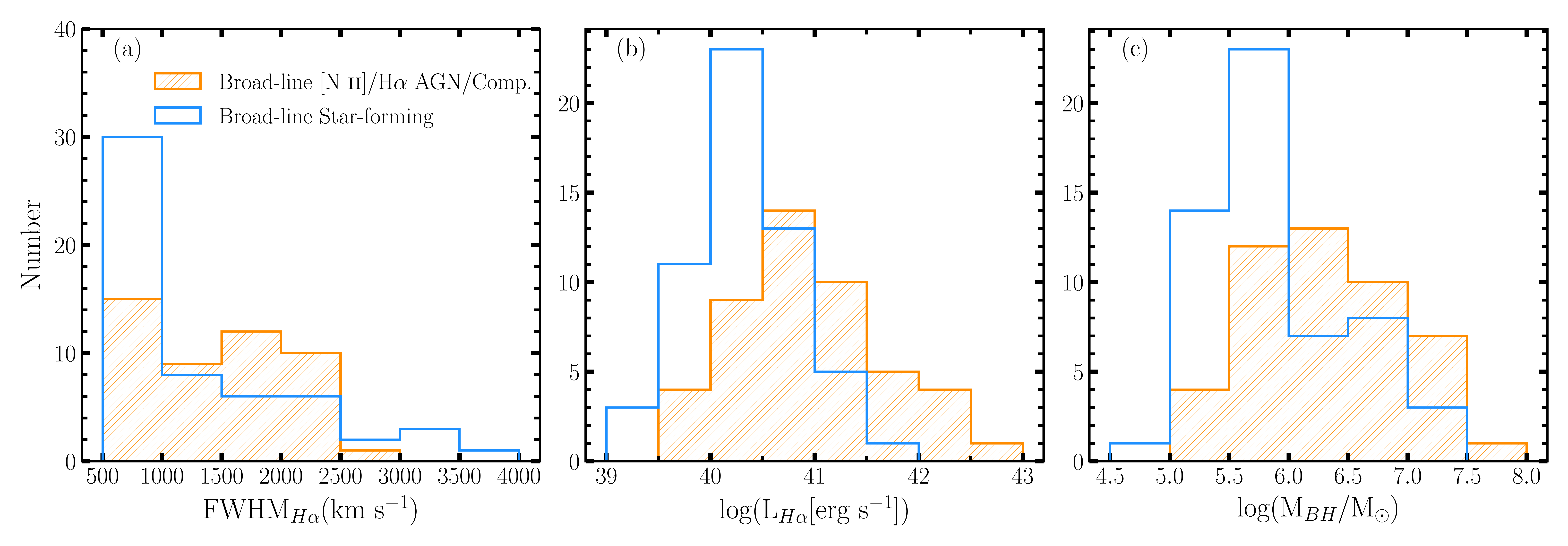}
\caption{Broad H$\alpha$ emission. \textit{Panels (a)--(b)}: The distributions of FWHM and log luminosity of the broad H$\alpha$ components. The [\ion{N}{2}]/H$\alpha$ AGN and composite galaxies are shown as hashed orange histograms, and the star-forming galaxies as blue histograms. See section \ref{sec:broadha} for details. \textit{Panel (c)}: The virial BH mass distribution for the broad-line AGN candidates. We use equation 5 in \citet{Reines:2013} to estimate these BH masses. See section \ref{sec:broadha} for details. Note that we do not include the broad-line star-forming objects in our final sample of AGNs.
} 
\label{fig:ha_dist}
\end{figure*}

The broad H$\alpha$ luminosities of the broad-line [\ion{N}{2}]/H$\alpha$-selected AGN candidates range from 10$^{39.7}-10^{42.6}$, with a median luminosity of  $10^{40.9}$ erg s$^{-1}$. The SF galaxies have broad H$\alpha$ components with lower luminosities that span a range of 10$^{39.1-41.6}$, with a median luminosity of $10^{40.3}$ erg s$^{-1}$. Moreover, the widths (FWHMs) of all the broad H$\alpha$ components span a range of $\sim$500--3664, with median FWHM of 1490 km s$^{-1}$ for the [\ion{N}{2}]/H$\alpha$ AGNs/Composites and 895 km s$^{-1}$ for the SF galaxies.
The distributions of FWHM and luminosity of the broad H$\alpha$ components are plotted in panels (a) and (b) of Figure \ref{fig:ha_dist}. Given that the luminosities and FWHMs of the broad H$\alpha$ lines in the SF galaxies tend to be significantly lower than those of the [\ion{N}{2}]/H$\alpha$ AGNs/Composites, and that many star-forming galaxies with broad H$\alpha$ are not in fact AGNs \citep{Baldassare:2016}, we consider these objects suspect and do not include them in our final sample of AGNs.

We estimate virial BH masses for the 47 broad-line AGNs/Composites using equation 5 in \citet{Reines:2013} and our measurements of the luminosity and FWHM of the broad H$\alpha$ line.  The resulting BH masses vary from $10^{5-7.7}$, with a median BH mass of $10^{6.2} M_\odot$.  We plot the distribution of BH masses in panel (c) of Figure \ref{fig:ha_dist}. A list of luminosities and FWHMs of the broad H$\alpha$ components, and the corresponding BH masses, for the AGNs/Composites are given in Table \ref{tab:bh_masses}. For the sake of completeness, we also estimate BH masses for the SF galaxies with broad H$\alpha$.  These are in the range of $10^{4.9-7.3}$, with a median of $10^{5.8}$.

The BH masses for the rest of the AGN candidates in this work are unknown. However, if we assume the BH mass-to-total stellar mass relation for AGNs derived in \citet{Reines:2015}, the BH masses for all of the AGN candidates span a range of 10$^{4.3}-$10$^{6.4}$, with a median BH mass of 10$^{6.2}$ M$_\odot$.

\begin{deluxetable}{cccc}
\tabletypesize{\scriptsize}
\tablecaption{Sample of Broad-line AGNs}
\tablehead{
\colhead{CATAID}&\colhead{log $L(H\alpha)_b$} & \colhead{FWHM(H$\alpha$)$_b$}&\colhead{log $M_{BH}$}
}
\decimalcolnumbers
\startdata
\multicolumn{ 4}{c}{[\ion{N}{2}]/H$\alpha$ AGNs} \\ \hline
2258819 & 41.44 &      926 &   6.2 \\
1787285 & 41.35 &      889 &   6.2 \\
346048 & 40.04 &      822 &   5.5 \\
382771 & 40.70 &      690 &   5.6 \\
656596 & 40.71 &      723 &   5.7 \\
3856528 & 41.47 &     2247 &   7.0 \\
3578870 & 41.17 &     1041 &   6.2 \\
852508 & 41.29 &      994 &   6.2 \\
689919 & 41.85 &     2160 &   7.2 \\
273195 & 40.94 &     1205 &   6.2 \\
537437 & 40.61 &     1296 &   6.1 \\
703117 & 41.88 &     1778 &   7.0 \\
696560 & 42.02 &     2118 &   7.3 \\
544030 & 40.88 &     1395 &   6.3 \\
238411 & 39.85 &     1582 &   6.0 \\
320888 & 40.98 &     2352 &   6.9 \\
740319 & 41.07 &     2043 &   6.8 \\
484908 & 40.80 &      574 &   5.5 \\
492449 & 41.67 &     1786 &   6.9 \\
48050 & 40.07 &     1628 &   6.1 \\
5247018 & 42.40 &     1642 &   7.2 \\
5240292 & 42.46 &     2206 &   7.5 \\
5265117 & 40.09 &     1442 &   6.0 \\
5275222 & 42.63 &     1543 &   7.3 \\
5220386 & 42.48 &     2670 &   7.7 \\
\hline
\multicolumn{ 4}{c}{[\ion{N}{2}]/H$\alpha$ Composites} \\ \hline
1125779 & 41.15 &     1620 &   6.6 \\
2008726 & 40.88 &      699 &   5.7 \\
1672767 & 40.74 &     2100 &   6.6 \\
1791657 & 40.43 &      962 &   5.8 \\
2123678 & 39.72 &      592 &   5.0 \\
375406 & 40.84 &      833 &   5.9 \\
3901665 & 41.53 &     2055 &   7.0 \\
727091 & 41.32 &     2097 &   6.9 \\
521922 & 39.81 &     1656 &   6.0 \\
609225 & 40.13 &     1463 &   6.0 \\
31941 & 39.74 &      721 &   5.2 \\
98560 & 40.86 &      707 &   5.7 \\
718498 & 40.94 &     2355 &   6.8 \\
296972 & 40.54 &     1490 &   6.2 \\
297773 & 40.13 &     1211 &   5.9 \\
297764 & 40.24 &      559 &   5.2 \\
593724 & 41.37 &     1620 &   6.7 \\
267078 & 40.25 &     1550 &   6.1 \\
62983 & 40.27 &     1618 &   6.2 \\
5197149 & 41.58 &     1681 &   6.8 \\
5337331 & 41.00 &     1135 &   6.2 \\
5266552 & 41.29 &      742 &   6.0 \\
\enddata
\tablecomments{Column 1: Unique ID of the GAMA object. Column 2: The luminosity of the broad H$\alpha$ component in units of erg s$^{-1}$. Column 3: The width (FWHM) of the broad H$\alpha$ component in units of km s$^{-1}$, corrected for instrumental resolution. Column 4: The virial mass estimate of the BH in units of M$_\odot$ by assuming the broad H$\alpha$ emission is associated with the BLR. See section \ref{sec:broadha} for more details.}
\label{tab:bh_masses}
\end{deluxetable}


\section{Sample Properties}
\label{sec:host_properties}

\subsection{Newly-Identified AGNs and Active Fractions}
\label{sec:syn_res}

In this work, we identify 388 unique AGN candidates from our parent sample of low-mass galaxies by utilizing two narrow-line diagnostic diagrams (sections \ref{sec:nii} and \ref{sec:he2}) as well as searching for [\ion{Fe}{10}]$\lambda$6374 and [\ion{Ne}{5}]$\lambda3426$ coronal-line emission (sections \ref{sec:fex} and \ref{sec:ne5}). 
We do not find any matches between our parent sample of galaxies and the AGNs reported in \citet{Greene:2007}, \citet{Reines:2013}, \citet{Moran:2014}, \citet{Chilingarian:2018}, and \citet{Molinafex:2021}. In fact, only 2164/23460 galaxies in our parent sample have been observed by other surveys, out of which only 301 have SDSS spectra. Thus, we conclude that this work presents an entirely new sample of AGNs in low-mass galaxies.

Overall we find an active fraction among our parent sample of low-mass emission-line galaxies of $388/23460 \approx 1.7\%$. Accounting for all of the low-mass galaxies in GAMA (including those that were cut from our parent sample due to weak/no lines, see \S\ref{sec:sample_selection}), the active fraction drops to $388/52782 \approx 0.7\%$. The majority of the active galaxies were found as AGNs/composites in the [\ion{N}{2}]/H$\alpha$ diagnostic diagram. These alone give an active fraction of $\sim$ 1.3\% among our parent sample of low-mass emission-line galaxies. The active fraction using the \ion{He}{2}/H$\beta$ ratio and [\ion{Fe}{10}]-emitting galaxies are each $\sim$ 0.2\%, and the fraction of detectable [\ion{Ne}{5}]-emitting galaxies is just $\sim$ 0.01\%. 
While accurate comparisons to other spectroscopic searches for AGNs in the low-mass regime are complicated by various selection criteria and the differing survey characteristics, the values we find are in approximate agreement with prior work \citep{Reines:2013,Moran:2014,Sartori:2015,Molinafex:2021,Polimera:2022}.


\subsection{Host Galaxies}
\label{sec:host_galaxies}

\begin{figure*}[bhtp]
\centering
\includegraphics[width=\textwidth]{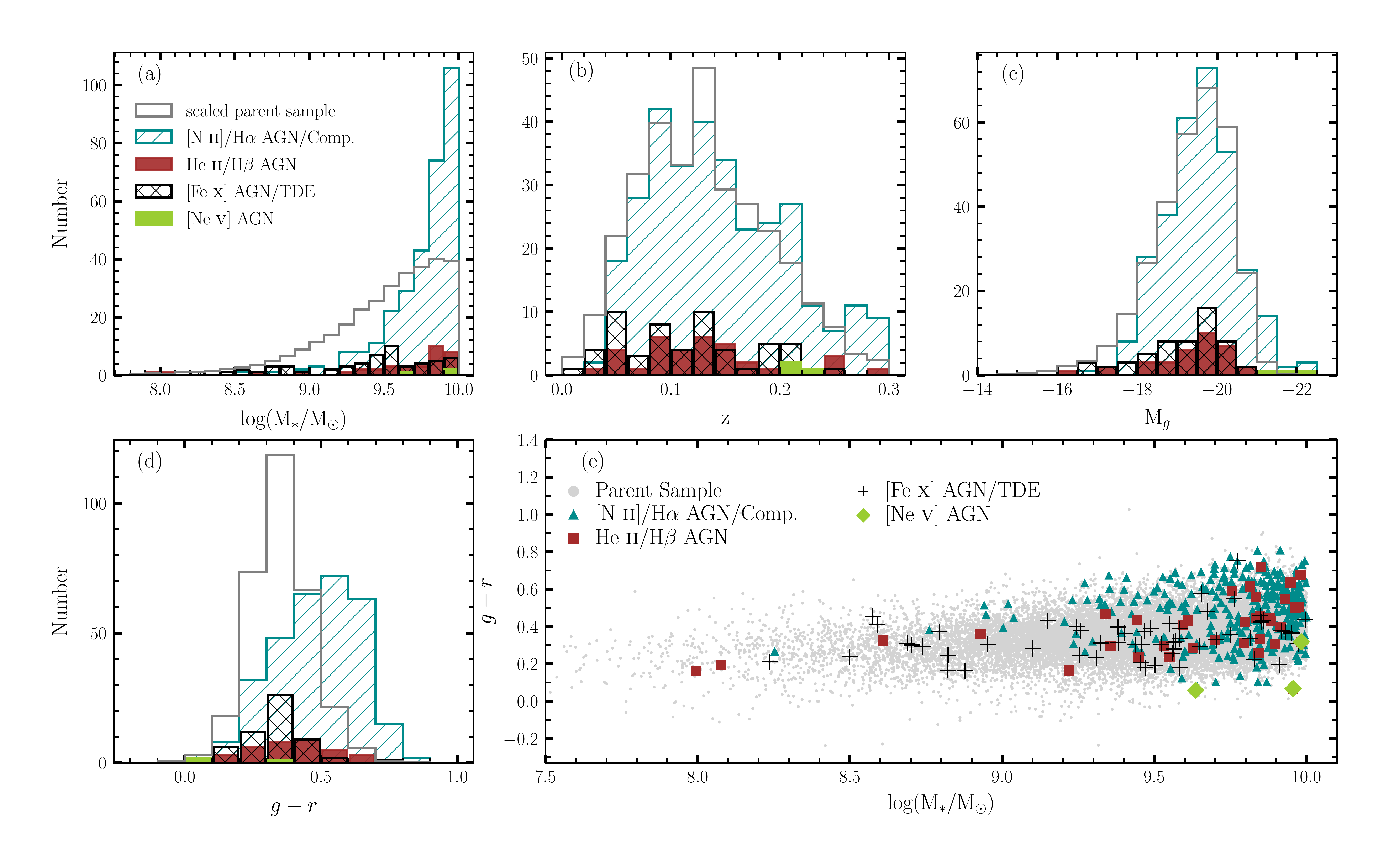}
\caption{Host galaxy properties. \textit{(a)--(d)}: Distributions of host galaxy stellar mass, redshift, absolute $g$-band magnitude and $g-r$ color.
The [\ion{N}{2}]/H$\alpha$-selected AGN candidates are shown as hashed teal histograms, the \ion{He}{2}/H$\beta$-selected AGNs are plotted as brown histograms, the [\ion{Fe}{10}] AGNs galaxies are shown as cross-hashed black histograms, and the [\ion{Ne}{5}]-AGNs are shown as green histograms. Our parent sample of low-mass galaxies is also shown as gray histograms, normalized to the number of [\ion{N}{2}]/H$\alpha$-selected AGN and composite galaxies. 
All values are adopted from \texttt{StellarMassesG02CFHTLS v24} and \texttt{StellarMassesGKV v24} tables \citep{Taylor:2011,Bellstedt:2020}. 
\textit{(e)}: The $g-r$ vs.\ log(M$_*/$M$_\odot$) plot. Here we use similar color scheme as panels (a)--(d). The [\ion{N}{2}]/H$\alpha$ AGNs/Composites are among redder and more massive objects in our parent sample, which are similar to \ion{He}{2}/H$\beta$ AGNs. The [\ion{Fe}{10}] AGNs seem to trace the general trend of our parent sample, while the [\ion{Ne}{5}] AGNs are among more massive and bluer objects.
}
\label{fig:distributions}
\end{figure*}

The host galaxies of the AGNs in our sample have an upper mass limit of $10^{10}$ M$_\odot$ by design, and the lowest-mass galaxies with AGNs in our sample 
have stellar masses of log$(M_*/M_\odot) \sim 8$ (see Figure \ref{fig:distributions}). A summary of the host galaxy properties for each sub-sample can be found in Table \ref{tab:other_bh_mass} (also see Figure \ref{fig:distributions} and Table \ref{tab:gal_prop} for individual values). Consistent with previous studies \citep[e.g.,][]{Reines:2013}, the [\ion{N}{2}]/H$\alpha$-selected AGNs are predominantly among the higher-mass galaxies, although the minimum galaxy mass in this sub-sample has log$(M_*/M_\odot) \sim 8.3$. The \ion{He}{2}-selected AGNs show a similar trend.
In contrast, the [\ion{Fe}{10}]-emitting galaxies are more evenly spread out in terms of their stellar mass and tend to be more reflective of the parent sample of low-mass galaxies. The rare [\ion{Ne}{5}]-emitting galaxies are exclusively found among higher mass objects with luminous AGNs.

The median total absolute $g$-band magnitude of our sample is 
$\langle M_g \rangle=-19.5$ mag. 
This is very similar to that of the \citet{Greene:2007} sample of broad-line AGNs with BH masses $M_{\rm BH} \lesssim 2 \times 10^6~M_\odot$.
Our median $g$-band magnitude is also $\sim$1 mag more luminous than what \citet{Reines:2013} found for their AGN sample and that of LMC \citep[M$_g^{LMC} \sim -18.2$ mag; ][]{Tollerud:2011}. This is not surprising given that our upper mass limit is more than 3 times larger than that in the \citet{Reines:2013} sample.  
 
A color-mass diagram is 
shown in panel (e) of Figure \ref{fig:distributions}. The host galaxies of the [\ion{N}{2}]/H$\alpha$-selected AGN candidates tend to be redder and relatively massive overall, consistent with the findings in \citet{Reines:2013}. The bias towards redder galaxies selected using the [\ion{N}{2}]/H$\alpha$ diagnostic may be a selection effect, since this diagnostic is metallicity sensitive and struggles with detecting AGNs in low-metallicity star-forming galaxies \citep[e.g., ][]{Groves:2006,Reines:2013,Kewley:2019}. On the other hand, the galaxies among the \ion{He}{2}/H$\beta$ and [\ion{Fe}{10}] sub-samples  
extend to less massive and bluer, thus more star-forming galaxies.  \citet{Molinafex:2021} found a similar trend for [\ion{Fe}{10}]-emitting dwarf galaxies in the SDSS.  The [\ion{Ne}{5}]-emitting sub-sample are among galaxies that are more massive and bluer in our parent sample, characteristic of quasars with strong UV emission from the accretion disk. Given that these objects are powered by relatively low-mass BHs, they may be akin to ``miniquasars" that have been proposed as potential contributors to cosmic reionization \citep{Haiman:1998,Madau:2004}. 

The redshift distributions of the active galaxies in this work, along with that of our parent sample of low-mass galaxies, are shown in panel (b) of Figure \ref{fig:distributions}. The maximum redshift of $z=0.3$ comes from our requirement of detecting and modeling the narrow-line profile using the [\ion{S}{2}]$\lambda\lambda$6716,6731 doublet (\S\ref{sec:lines}). 
Overall, the median redshift of the active galaxies is $z = 0.13$. The [\ion{Ne}{5}] line is only within the observable wavelength range for $z\geq0.15$ and therefore the three [\ion{Ne}{5}]-AGNs are at higher redshifts than the other sub-samples (see Table \ref{tab:other_bh_mass}). The [\ion{Fe}{10}]-selected AGNs/TDEs are at slightly lower redshifts compared to the [\ion{N}{2}]/H$\alpha$ and \ion{He}{2}-selected objects, likely owing to the weakness of the [\ion{Fe}{10}] line.   

\begin{deluxetable*}{lcccccccccccc}

\tablecaption{Host Galaxy Properties}
\tabletypesize{\scriptsize}
\setlength{\tabcolsep}{4pt}
\renewcommand{\arraystretch}{1.}
\tablewidth{2pt}
\tablehead{ 
\vspace{-2mm}\\
\colhead{}&\multicolumn{3}{c}{log(M$_*$/M$_\odot$)}&\multicolumn{3}{c}{z}&\multicolumn{3}{c}{M$_g$}&\multicolumn{3}{c}{$g-r$}\vspace{-2mm}\\
\multicolumn{1}{c}{}&\multicolumn{3}{c}{\hrulefill}&\multicolumn{3}{c}{\hrulefill}&\multicolumn{3}{c}{\hrulefill}&\multicolumn{3}{c}{\hrulefill}\\
\colhead{Diagnostic}&\colhead{min.}&\colhead{max.}&\colhead{med.}&\colhead{min.}&\colhead{max.}&\colhead{med.}& \colhead{min.}&\colhead{max.}&\colhead{med.}& \colhead{min.}&\colhead{max.}&\colhead{med.} \\
{}&(1)&(2)&(3)&(4)&(5)&(6)&(7)&(8)&(9)&(10)&(11)&(12)\\
}
\startdata
[\ion{N}{2}]/H$\alpha$ &      8.25 &     10.00 &      9.84 & 0.0277 & 0.2974 & 0.1365 &   -16.58 &   -22.25 &   -19.64 &     0.06 &     0.81 &     0.50 \\
\hline
\ion{He}{2}/H$\beta$ &      7.99 &      9.98 &      9.81 & 0.0289 & 0.2871 & 0.1289 &   -16.40 &   -21.81 &   -19.64 &     0.07 &     0.72 &     0.40 \\
\hline
[\ion{Fe}{10}]$\lambda$6374 &      8.24 &     10.00 &      9.50 & 0.0132 & 0.2421 & 0.1072 &   -16.51 &   -21.30 &   -19.45 &     0.16 &     0.75 &     0.32 \\
\hline
[\ion{Ne}{5}]$\lambda$3426 &      9.64 &      9.98 &      9.96 & 0.2095 & 0.2348 & 0.2102 &   -21.09 &   -22.07 &   -21.81 &     0.06 &     0.32 &     0.07 \\
\enddata
\tablecomments{Summary of the host galaxy properties of the low-mass AGNs using the four narrow-line diagnostics applied in this work. We show the minimum, maximum, and median values of the host galaxy stellar mass, redshift, $g$-band absolute magnitude, and $g-r$ color for each of the diagnostics. All the values are adopted from the \texttt{StellarMassesG02CFHTLS v24} and \texttt{StellarMassesGKV v24} tables \citep{Taylor:2011,Bellstedt:2020}. 
We exclude one of the [\ion{N}{2}]/H$\alpha$-selected candidates (CATAID 5227891) when calculating these values because it has an erroneous M$_g$ estimate. 
See section \ref{sec:host_galaxies} for more details.
}
\label{tab:other_bh_mass}
\end{deluxetable*}

Our sample of active galaxies extends to higher redshifts than previous samples in the low-mass regime based on SDSS spectroscopy.  For example, the \citet{Reines:2013} dwarf galaxies all have $z\lesssim0.055$, the \citet{Moran:2014} sample has $z\lesssim0.018$, \citet{Sartori:2015} finds a median redshift of ${z}\sim0.03$, and the [\ion{Fe}{10}]-selected objects in \citet{Molinafex:2021} have a median redshift of $z\sim 0.03$.
The closest comparisons are to that of the Type 1 AGN sample of \citet{Greene:2007}, which has a median redshift of 0.08, and the Type 2 AGN counterparts in \cite{Barth:2008} that have $z\lesssim0.08$.
The higher redshifts probed by our study are likely due to the fact that the GAMA spectroscopic limiting magnitude is $\sim 2$ magnitudes deeper than that of the SDSS.


\subsection{The Dwarf Galaxy Sample}
\label{sec:dwarfs}

Searches for AGNs in the low-mass regime often use different criteria for selecting their samples. In some cases, low BH masses are used \citep[e.g.,][]{Greene:2007,Chilingarian:2018} and in others, absolute magnitude \citep{Barth:2008} or stellar mass limits \citep[e.g.,][]{Reines:2013} are used. As described above, our main sample of low-mass active galaxies has an upper stellar mass limit of $10^{10} M_\odot$, which has also been used by \citet{Moran:2014} and \citet{Baldassare:2018}. Here we focus on AGNs in the dwarf galaxy mass range, which is usually taken to be $M_\star \leq 3\times10^9 M_\odot$ \citep{Reines:2013}.

\begin{figure*}[htb!]
\centering
\includegraphics[width=1\textwidth]{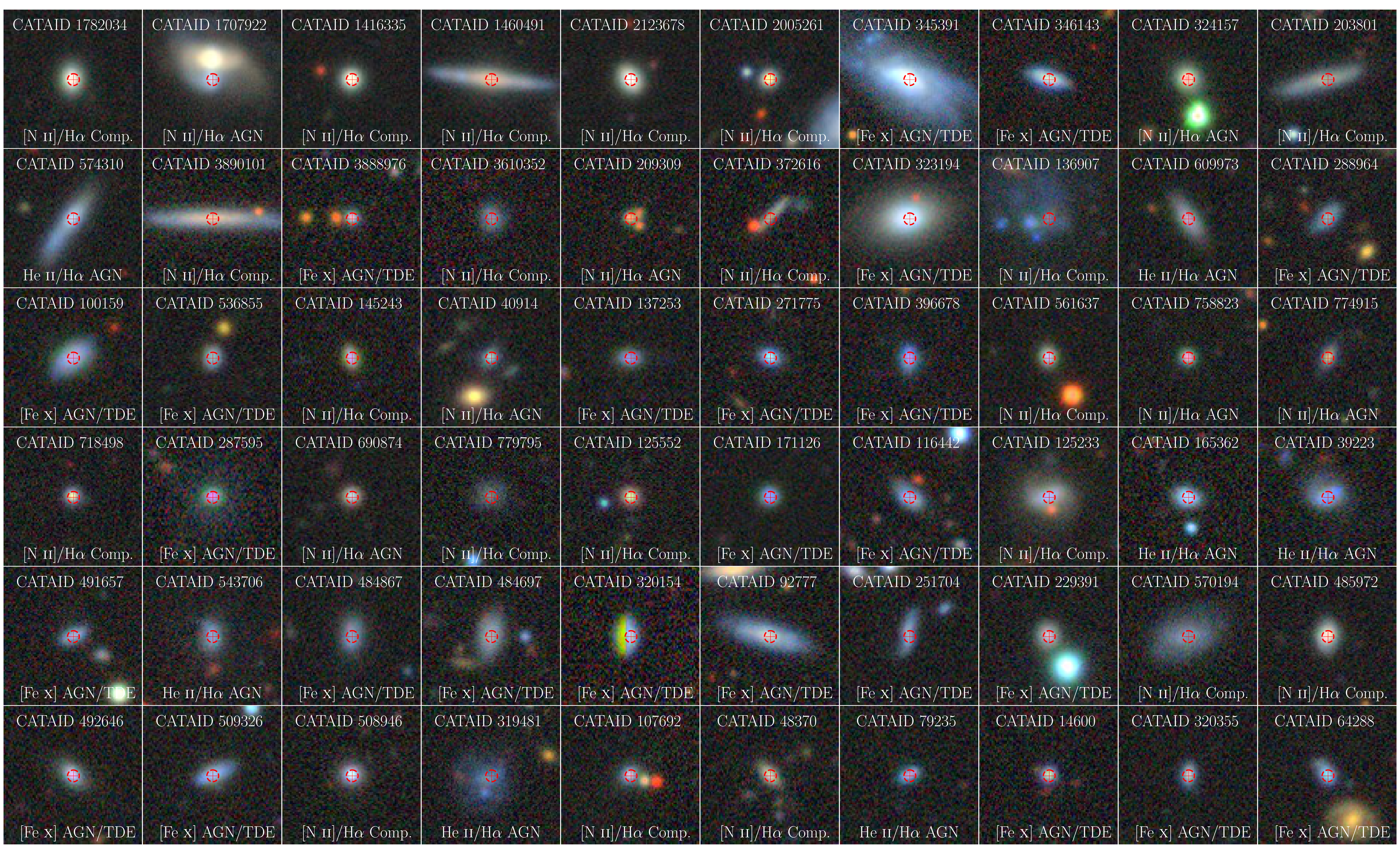}
\caption{DESI Legacy Imaging Survey SkyViewer $grz-$band images of dwarf galaxies ($M_\star \leq 3\times10^9 M_\odot$) in our sample with available images (60/70). Red crosses indicate the positions of the spectra and the red dashed circles indicate the sizes of the GAMA spectroscopic fibers (with 2$\arcsec$ diameters). }
\label{fig:dwarf_images}
\end{figure*}

As discussed in Section \ref{sec:data}, our parent sample of low-mass emission-line galaxies with $M_\star \leq 10^{10} M_\odot$ consists of 23,460 objects, of which 9,094 are dwarf galaxies with $M_\star \leq 3\times10^9 M_\odot$. In total, we identify 70 unique dwarf galaxies hosting AGNs based on our diagnostics described in Section \ref{sec:results}. We find 9 AGNs and 25 composites using the [\ion{N}{2}]/H$\alpha$ diagram. Two of the dwarf composites also have broad H$\alpha$ emission and virial BH masses of $\sim 10^5 M_\odot$ and $\sim 7 \times 10^6 M_\odot$. 
There are 13 dwarf galaxies with detectable \ion{He}{2} emission, 9 of which are AGN candidates with high \ion{He}{2}/H$\beta$ ratios.  
We find that 27 of the [\ion{Fe}{10}]-emitting galaxies are dwarf galaxies, while none of the [\ion{Ne}{5}]-emitting galaxies are in this mass range. We show $grz-$band images of most of the dwarf galaxies in our sample in Figure \ref{fig:dwarf_images}, which we obtained from the DESI Legacy Imaging Survey SkyViewer \citep{decals}. 

\section{Summary and Conclusions}
\label{sec:discussion_summary}

In this work, we have systemically searched for optical signatures of active massive BHs in $\sim$23,000 galaxies with stellar masses $M_\star\leq10^{10} M_\odot$ and redshifts $z\leq 0.3$ by analyzing spectroscopic data from GAMA DR4.
We employed four optical emission-line diagnostics and identified 388 unique active galaxies, 70 of which are in the dwarf galaxy regime with $10^8 \lesssim M_\star/M_\odot \lesssim 10^{9.5}$. Our main results are summarized in Figures \ref{fig:AGNsamp} and \ref{fig:distributions}.
 
We used the ratio of [\ion{O}{3}]/H$\beta$ vs.\ [\ion{N}{2}]/H$\alpha$ as our first diagnostic. This diagnostic diagram has previously been used to identify AGNs in low-mass/dwarf galaxies \citep[e.g., ][]{Reines:2013,Moran:2014,Sartori:2015}, and follow-up observations with X-rays have confirmed the existence of massive BHs in some of these sources independently \citep[e.g., ][]{Baldassare:2017}.
Moreover, the clean separation between the AGN and composite galaxies, and those consistent with star formation make these AGN candidates easily distinguishable. For these reasons, we consider the 71 AGNs identified by this diagnostic as secure and the 238 composite galaxies as strong AGN candidates. While this diagnostic provides a relatively clean sample, it can miss weakly accreting BHs and/or those residing in actively star-forming galaxies (particularly those with low metallicities).

Next, we searched for low-mass galaxies with relatively high \ion{He}{2}/H$\beta$ ratios. 
We employed a stricter criterion, log(\ion{He}{2}/H$\beta) > -1$, to select AGNs than previous works \citep[e.g., ][]{Shirazi:2012,Sartori:2015} with the goal of providing a clean sample. This ratio is expected to be higher than what can be produced by stellar-mass X-ray binaries and Wolf-Rayet stars 
\citep{Schaerer:2019}. We find 36 galaxies that meet this criterion. Of these, 10 are also [\ion{N}{2}]/H$\alpha$ AGNs and 1 is a composite. Given that the majority of the AGN candidates identified by this diagnostic are star-forming in the [\ion{N}{2}]/H$\alpha$ diagram, further observations are needed to confirm our results independently. 

We also systematically searched for two high-ionization coronal lines ([\ion{Fe}{10}]$\lambda6374$ and [\ion{Ne}{5}]$\lambda3426$) in the spectra of our parent sample of low-mass galaxies.  The [\ion{Fe}{10}]$\lambda6374$ coronal line is detectable in 56 galaxies, only 2 of which have additional AGN indicators. As discussed in detail in \citet{Molinafex:2021}, [\ion{Fe}{10}]$\lambda6374$ can be produced by certain types of supernovae. However, one of the most extreme known examples is SN 2005ip supernovae with a peak luminosity of $2 \times 10^{37}$ erg s$^{-1}$ \citep{smith:2009}. This is an order of magnitude less than the minimum [\ion{Fe}{10}] luminosity of $10^{38}$ erg s$^{-1}$ in our sample. Thus, we are optimistic that these [\ion{Fe}{10}] lines are produced by AGN activity or extreme coronal-line emitting TDEs, both of which require massive BHs. We found three galaxies with strong [\ion{Ne}{5}] emission that are also [\ion{N}{2}]/H$\alpha$ AGNs. Two of these objects were also selected as AGNs using our \ion{He}{2}/H$\beta$ criterion.

In total we have found 388 unique low-mass galaxies exhibiting narrow-line signatures of active massive BHs, 47 of which have detectable broad H$\alpha$ emission in their spectra. Using standard virial techniques, we estimated BH masses for these objects and find a range of $M_{\rm BH} \sim 10^{5.0-7.7} M_\odot$. The median BH mass is $10^{6.2} M_\odot$, consistent with expectations given the host galaxy stellar masses \citep{Reines:2015}. We found an additional 56 star-forming galaxies with broad H$\alpha$ emission in their spectra, with no narrow-line signatures indicating the presence of AGNs. Given that broad H$\alpha$ in many star-forming dwarf galaxies can be produced by transient stellar processes such as supernovae \citep{Baldassare:2016}, 
we are suspicious of the broad-line objects without narrow-line signatures of AGNs and do not include them in our final sample of low-mass active galaxies.   

As seen in previous works \citep{Reines:2013,Molinafex:2021}, the various emission-line AGN diagnostics that we have used tend to probe different parts of the parameter space spanned by our parent sample of low-mass/dwarf galaxies (see Figure \ref{fig:distributions}). For example, the [\ion{N}{2}]/H$\alpha$ AGNs/Composites are biased towards redder and more massive galaxies within our parent sample, and the [\ion{Fe}{10}]-selected AGNs tend to be in bluer star-forming galaxies with a color and mass distribution more representative of our parent sample. Thus, using a multi-diagnostic approach can provide a more complete census of AGNs in low-mass/dwarf galaxies. While we have strived to strike a balance between assembling a clean yet comprehensive sample of low-mass/dwarf active galaxies in GAMA, large-scale follow-up campaigns would be useful to check the robustness of the AGN diagnostics we (and others) have applied in the low-mass regime. 

Ultimately this work has provided an entirely new sample of hundreds of low-mass/dwarf active galaxies, which extends to southern sky regions and higher redshifts than previous searches in the low-mass regime. We find an AGN fraction of $\sim 1\%$, which is similar to other spectroscopic searches in this mass range. This active fraction provides a lower limit on the BH occupation fraction in low-mass galaxies with implications for the origin of the first BH seeds.


\acknowledgements
We thank the anonymous reviewer for their helpful comments and suggestions that improved this work. AER acknowledges support for this work provided by Montana State University and NASA through EPSCoR grant number 80NSSC20M0231.

MM is supported by funding from Ford Foundation Postdoctoral Fellowship, administered by the National Academies of Sciences, Engineering, and Medicine, awarded to MM in 2021-2022.

GAMA is a joint European-Australasian project based around a spectroscopic campaign using the Anglo-Australian Telescope. The GAMA input catalogue is based on data taken from the Sloan Digital Sky Survey and the UKIRT Infrared Deep Sky Survey. Complementary imaging of the GAMA regions is being obtained by a number of independent survey programmes including GALEX MIS, VST KiDS, VISTA VIKING, WISE, Herschel-ATLAS, GMRT and ASKAP providing UV to radio coverage. GAMA is funded by the STFC (UK), the ARC (Australia), the AAO, and the participating institutions. The GAMA website is http://www.gama-survey.org/. Based on observations made with ESO Telescopes at the La Silla Paranal Observatory under programme ID 179.A-2004. Based on observations made with ESO Telescopes at the La Silla Paranal Observatory under programme ID 177.A-3016.

The Legacy Surveys consist of three individual and complementary projects: the Dark Energy Camera Legacy Survey (DECaLS; Proposal ID 2014B-0404; PIs: David Schlegel and Arjun Dey), the Beijing-Arizona Sky Survey (BASS; NOAO Prop. ID 2015A-0801; PIs: Zhou Xu and Xiaohui Fan), and the Mayall z-band Legacy Survey (MzLS; Prop. ID 2016A-0453; PI: Arjun Dey). DECaLS, BASS and MzLS together include data obtained, respectively, at the Blanco telescope, Cerro Tololo Inter-American Observatory, NSF’s NOIRLab; the Bok telescope, Steward Observatory, University of Arizona; and the Mayall telescope, Kitt Peak National Observatory, NOIRLab. The Legacy Surveys project is honored to be permitted to conduct astronomical research on Iolkam Du’ag (Kitt Peak), a mountain with particular significance to the Tohono O’odham Nation.

NOIRLab is operated by the Association of Universities for Research in Astronomy (AURA) under a cooperative agreement with the National Science Foundation.

This project used data obtained with the Dark Energy Camera (DECam), which was constructed by the Dark Energy Survey (DES) collaboration. Funding for the DES Projects has been provided by the U.S. Department of Energy, the U.S. National Science Foundation, the Ministry of Science and Education of Spain, the Science and Technology Facilities Council of the United Kingdom, the Higher Education Funding Council for England, the National Center for Supercomputing Applications at the University of Illinois at Urbana-Champaign, the Kavli Institute of Cosmological Physics at the University of Chicago, Center for Cosmology and Astro-Particle Physics at the Ohio State University, the Mitchell Institute for Fundamental Physics and Astronomy at Texas A\&M University, Financiadora de Estudos e Projetos, Fundacao Carlos Chagas Filho de Amparo, Financiadora de Estudos e Projetos, Fundacao Carlos Chagas Filho de Amparo a Pesquisa do Estado do Rio de Janeiro, Conselho Nacional de Desenvolvimento Cientifico e Tecnologico and the Ministerio da Ciencia, Tecnologia e Inovacao, the Deutsche Forschungsgemeinschaft and the Collaborating Institutions in the Dark Energy Survey. The Collaborating Institutions are Argonne National Laboratory, the University of California at Santa Cruz, the University of Cambridge, Centro de Investigaciones Energeticas, Medioambientales y Tecnologicas-Madrid, the University of Chicago, University College London, the DES-Brazil Consortium, the University of Edinburgh, the Eidgenossische Technische Hochschule (ETH) Zurich, Fermi National Accelerator Laboratory, the University of Illinois at Urbana-Champaign, the Institut de Ciencies de l’Espai (IEEC/CSIC), the Institut de Fisica d’Altes Energies, Lawrence Berkeley National Laboratory, the Ludwig Maximilians Universitat Munchen and the associated Excellence Cluster Universe, the University of Michigan, NSF’s NOIRLab, the University of Nottingham, the Ohio State University, the University of Pennsylvania, the University of Portsmouth, SLAC National Accelerator Laboratory, Stanford University, the University of Sussex, and Texas A\&M University.

BASS is a key project of the Telescope Access Program (TAP), which has been funded by the National Astronomical Observatories of China, the Chinese Academy of Sciences (the Strategic Priority Research Program “The Emergence of Cosmological Structures” Grant No. XDB09000000), and the Special Fund for Astronomy from the Ministry of Finance. The BASS is also supported by the External Cooperation Program of Chinese Academy of Sciences (Grant No. 114A11KYSB20160057), and Chinese National Natural Science Foundation (Grant No. 11433005).

The Legacy Survey team makes use of data products from the Near-Earth Object Wide-field Infrared Survey Explorer (NEOWISE), which is a project of the Jet Propulsion Laboratory/California Institute of Technology. NEOWISE is funded by the National Aeronautics and Space Administration.

The Legacy Surveys imaging of the DESI footprint is supported by the Director, Office of Science, Office of High Energy Physics of the U.S. Department of Energy under Contract No. DE-AC02-05CH1123, by the National Energy Research Scientific Computing Center, a DOE Office of Science User Facility under the same contract; and by the U.S. National Science Foundation, Division of Astronomical Sciences under Contract No. AST-0950945 to NOAO.
acknowledgements


\software{
Astropy \citep{astropy2013,astropy2018},
Matplotlib \citep{matplotlib},
LMFIT\citep{lmfit}}


\clearpage
\appendix
\section{Observed Spectra and Emission-Line Fits for the \ion{He}{2} AGNs and [\ion{Fe}{10}] AGNs/TDEs}
\label{appendix:lines}

\begin{figure*}[htbp]
\centering
\includegraphics[width=\textwidth]{figures/heii_spectra.pdf}
\caption{Redshift-corrected spectra of a selection of \ion{He}{2}/H$\beta$ AGNs overplotted with the continuum model (sky blue). These \ion{He}{2}/H$\beta$ AGNs are SF in the [\ion{N}{2}]/H$\alpha$ diagram.}
\label{fig:heii_spectra}
\end{figure*}

\begin{figure*}[htbp]
\centering
\includegraphics[width=0.99\textwidth]{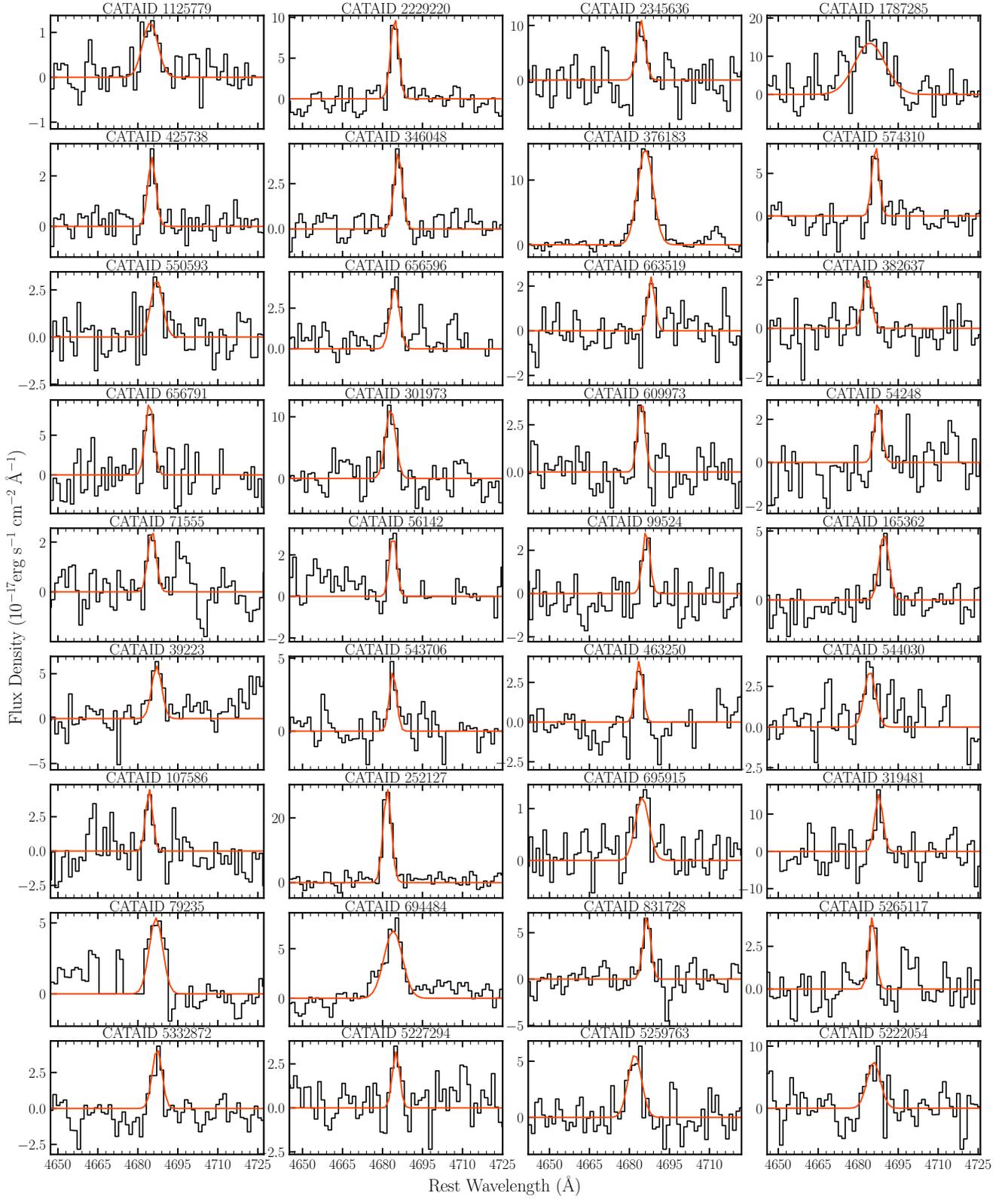}
\caption{Chunks of the \ion{He}{2} spectra and the single Gaussian emission-line fits for the 36 \ion{He}{2}/H$\beta$ AGN candidates. Here the continuum-subtracted spectrum are shown as black and the individual Gaussian component as orange-red. 10 of these galaxies are [\ion{N}{2}]/H$\alpha$ AGNs, while 1 is a composite object. See section \ref{sec:he2} for details.}
\label{fig:spec_he2}
\end{figure*}


\begin{figure*}[htbp]
\centering
\includegraphics[width=\textwidth]{figures/fex_spectra.pdf}
\caption{Redshift-corrected spectra of a selection of [\ion{Fe}{10}] AGNs/TDEs overplotted with the continuum model (sky blue). These objects are SF in the [\ion{N}{2}]/H$\alpha$ diagram.}
\label{fig:fex_spectra}
\end{figure*}

\begin{figure*}[htbp]
\centering
\includegraphics[width=
\textwidth]{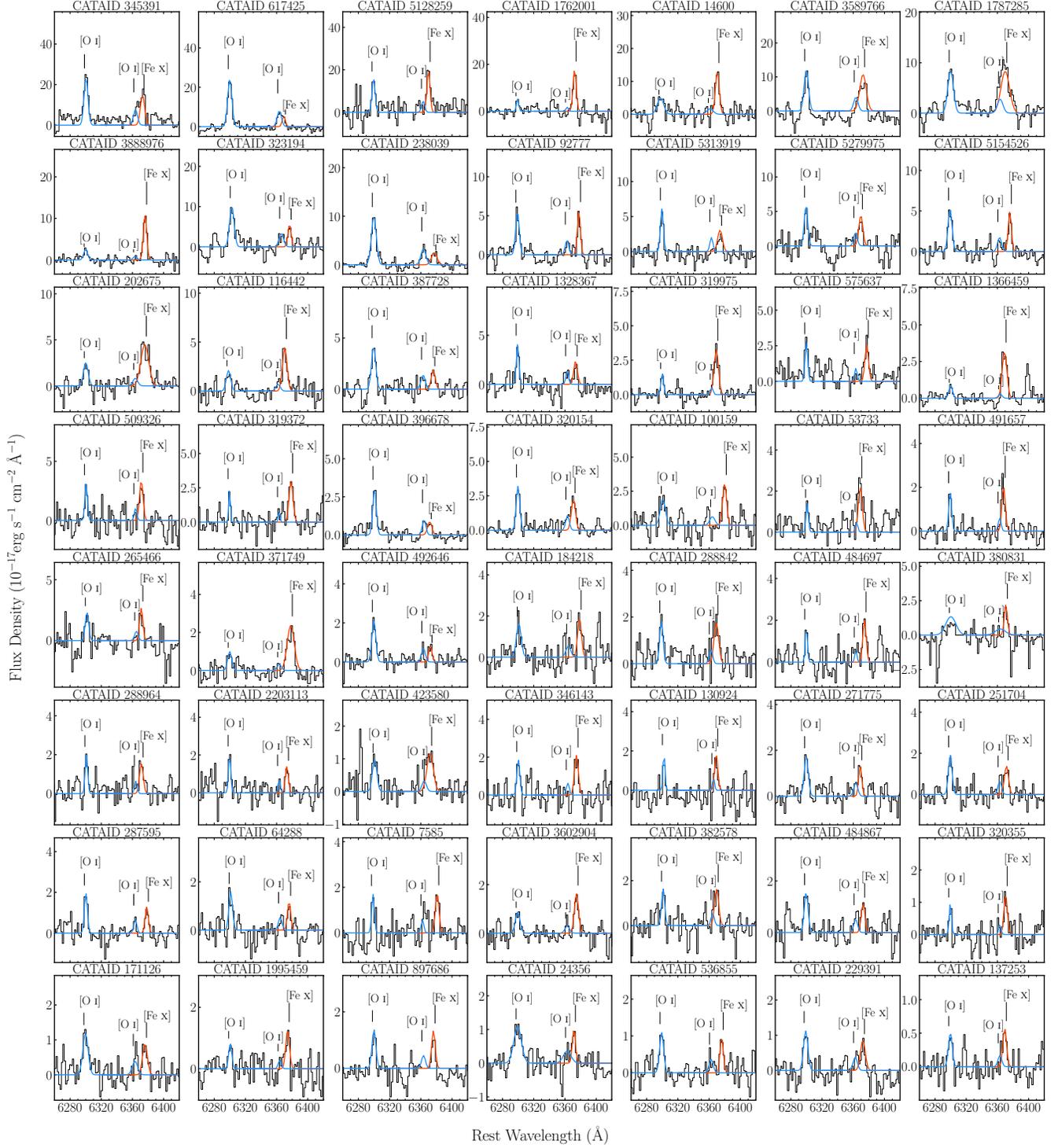}
\caption{Chunks of [\ion{Fe}{10}] emission-line spectra and the Gaussian emission-line fits for the 56 galaxies with this emission. We plot the single Gaussian models for the [\ion{O}{1}]$\lambda$6300,6363 lines in blue and [\ion{Fe}{10}]$\lambda$6374 line in red. Here the galaxy with CATAID 1787285 is an [\ion{N}{2}]/H$\alpha$ and \ion{He}{2}/H$\beta$ AGN, and the object with CATAID 387728 is an [\ion{N}{2}]/H$\alpha$ Composite. The [\ion{Fe}{10}] lines in our sample have luminosities that span a range of 10$^{38}-$10$^{41}$ erg s$^{-1}$, with a median of $10^{39.6}$ erg s$^{-1}$, which can be explained by AGNs or TDEs. See Section~\ref{sec:fex} for details. } \label{fig:oi_fex}
\end{figure*}

\clearpage
\bibliographystyle{aasjournal}
\bibliography{papers}

\end{document}